\documentclass[aps,
prc,
twocolumn,
nofootinbib,
noshowkeys,
noshowpacs,
10pt,
superscriptaddress,
floatfix]{revtex4-2}

\usepackage{graphicx}
\usepackage{float}
\usepackage{nicefrac}
\usepackage[normalem]{ulem}
\usepackage{amsmath,amssymb}
\usepackage{subfigure} 
\usepackage{mathtools}
\usepackage{enumerate}
\usepackage[colorlinks=true,linktocpage=true,linkcolor=blue,citecolor=red,urlcolor=blue,pdfencoding=unicode]{hyperref}

\renewcommand{\d}{\mathrm{d}}
\newcommand{\gt}{\tilde{\gamma}}
\newcommand{\dt}{\tilde{\partial}}
\newcommand{\sigt}{\tilde{\sigma}}
\newcommand{\Dt}{\tilde{D}}
\newcommand{\Rt}{\tilde{R}}
\newcommand{\ot}{\tilde{\omega}_0}
\newcommand{\oft}{\tilde{\omega}}
\newcommand{\otNS}{\tilde{\omega}_{0,\text{NS}}}
\newcommand{\kt}{\tilde{\kappa}_0}
\newcommand{\ktNS}{\tilde{\kappa}_{0,\text{NS}}}
\newcommand{\st}{\tilde{\mathfrak{t}}}
\newcommand{\stNS}{\tilde{\mathfrak{t}}_{\text{NS}}}
\newcommand{\vt}{\tilde{v}}
\newcommand{\ut}{\tilde{u}}
\newcommand{\dww}{\frac{\delta_{\omega\omega}}{\tau_\omega}}
\newcommand{\lwk}{\frac{\ell_{\omega\kappa}}{\tau_\omega}}
\newcommand{\aww}{\frac{\lambda_{\omega\omega}}{\tau_\omega}}
\newcommand{\awt}{\frac{\lambda_{\omega\mathfrak{t}}}{\tau_\omega}}
\newcommand{\atw}{\frac{\lambda_{\mathfrak{t}\omega}}{\tau_\mathfrak{t}}}
\newcommand{\dkk}{\frac{\delta_{\kappa\kappa}}{\tau_\kappa}}
\newcommand{\tkt}{\frac{\tau_{\kappa\mathfrak{t}}}{\tau_\kappa}}
\newcommand{\akk}{\frac{\lambda_{\kappa\kappa}}{\tau_\kappa}}
\newcommand{\dtt}{\frac{\delta_{\mathfrak{t}\mathfrak{t}}}{\tau_{\mathfrak{t}}}}
\newcommand{\att}{\frac{\lambda_{\mathfrak{t}\mathfrak{t}}}{\tau_{\mathfrak{t}}}}
\newcommand{\lkt}{\frac{\ell_{\kappa\mathfrak{t}}}{\tau_\kappa}}
\newcommand{\ltk}{\frac{\ell_{\mathfrak{t}\kappa}}{\tau_{\mathfrak{t}}}}
\newcommand{\ttw}{\frac{\tau_{\mathfrak{t}\omega}}{\tau_{\mathfrak{t}}}}
\newcommand{\tauw}{\tau_\omega}
\newcommand{\tauk}{\tau_\kappa}
\newcommand{\taut}{\tau_{\mathfrak{t}}}
\renewcommand{\t}{\mathfrak{t}}

\newcommand{\etas}{{\eta_s}}

\graphicspath{{./figs/}}

\begin{document}
	
\title{Spin Polarization of \texorpdfstring{$\boldsymbol{\Lambda}$}{\unichar{"039B}} hyperons from Dissipative Spin Hydrodynamics}
\author{Sapna} 
\email{sapna.snpth@gmail.com}
\affiliation{Department of Physics, Fakir Mohan University, Balasore 756019, India.}
\author{Sushant K. Singh} 
\email{sushant7557@gmail.com}
\affiliation{Department of Physics \& Astronomy, University of Florence, Via G. Sansone 1, 50019 Sesto Fiorentino, Florence, Italy}
\affiliation{Variable Energy Cyclotron Centre, 1/AF, Bidhan Nagar, Kolkata 700064, India}
\author{David Wagner} 
\email{david.wagner@unifi.it}
\email{dwagner@itp.uni-frankfurt.de}
\affiliation{Department of Physics \& Astronomy, University of Florence, Via G. Sansone 1, 50019 Sesto Fiorentino, Florence, Italy}
\affiliation{INFN Sezione di Firenze, Florence, Italy}
\date{\today} 
\bigskip

\begin{abstract}
We present a framework for spin dynamics in the quark-gluon plasma created in relativistic heavy-ion collisions. Under the approximation of small polarization, macroscopic spin degrees of freedom decouple from the background, and their evolution equations and transport coefficients have been computed using quantum kinetic theory of massive particles with nonlocal collisions. Employing this theory, we numerically solve dissipative relativistic spin hydrodynamics. We explore three interaction scenarios between constituent particles and apply this framework to compute both global and local spin polarization of $\Lambda$ hyperons in Au+Au collisions at $\sqrt{s_{NN}} = 200$ GeV. Our results show that the initially vanishing spin potential relaxes toward thermal vorticity, driving global polarization. Furthermore, we demonstrate that the sign of longitudinal polarization is sensitive to the interaction type, emphasizing the need for a consistent treatment of dissipative effects in spin hydrodynamics to describe experimental data.
\end{abstract}
	
\pacs{ }
	
\maketitle

\section{Introduction}
The study of spin polarization in relativistic heavy-ion collisions (HICs) has gained considerable interest in recent years, owing to the observation of global and local polarization of hyperons~\cite{STAR:2017ckg,STAR200_Pj,STAR:2019erd,ALICE:2019onw,STAR:2021beb,STAR2021:oxg,ALICE2022:ppz,STAR2023:sth}. Spin polarization observables provide probes of the vorticity and spin dynamics in the hot and dense fireball formed in relativistic HIC. The earliest attempt using a perturbative QCD-inspired model~\cite{Liang2005:fpp} to understand global polarization attributed it to the large orbital angular momentum, which polarizes the quarks in the same direction through spin-orbit coupling, analogous to the well-known Barnett effect~\cite{Barnett1935}. 

In the quantum-statistical approach, the spin degrees of freedom are assumed to be in local thermodynamic equilibrium at the fireball freezeout, making it possible to define a density operator and derive an analytical expression for spin polarization at leading order in terms of the thermal vorticity, $\varpi^{\mu\nu}=\frac12(\partial^{\nu} \beta^{\mu}-\partial^{\mu} \beta^{\nu})$, where $\beta^\mu=u^\mu/T$ is the inverse four-temperature, with $u^\mu$ denoting the fluid four-velocity and $T$ the temperature~\cite{Becattini:2015ska}. The spin polarization generated solely from thermal vorticity shows very good agreement with experimental data on the global polarization of $\Lambda$-hyperons across Beam Energy Scan collision energies~\cite{Karpenko:2016jyx}. However, the data on local polarization, particularly the longitudinal polarization, exhibits an opposite sign compared to hydrodynamic predictions. It was soon realized that the thermal shear, defined as $\xi^{\mu\nu}=\frac12(\partial^{\nu}\beta^{\mu}+\partial^\mu \beta^\nu)$, provides an additional contribution to spin polarization~\cite{Becattini:2021suc,Becattini:2021iol,Fu:2021pok,Liu:2021uhn,Palermo:2024tza,Hidaka:2018nrc,Cong:2021lsp}, with a sign opposite to that of thermal vorticity. The combined effect of thermal vorticity and thermal shear leads to a net negative longitudinal polarization. However, the correct sign and magnitude of the observed $\Lambda$ polarization in Au+Au collisions at $\sqrt{s_{NN}}=200$ GeV can be reproduced only under the isothermal approximation, where gradients of temperature on the freezeout surface are neglected~\cite{Becattini:2021iol}.

Although the approach taking the spin degrees of freedom to be in local equilibrium describes the data well, it assumes the timescales for spin relaxation to be vanishingly small. However, several studies utilizing microscopic interactions have demonstrated that the spin-relaxation timescale is comparable to the lifetime of the fireball in relativistic HIC~\cite{Kapusta2020:gfa,Ayala2020:sjd,Kapusta2020:sai,Hidaka:2023oze,Ayala2020:wdb,Hongo2022:adb,Wagner:2024fhf}. This necessitates a dynamical treatment of the spin degrees of freedom, where the so-called spin potential $\Omega^{\mu\nu}_0$, representing the thermodynamically conjugate variable to the total angular momentum, is not taken to be equal to the thermal vorticity, but remains a dynamical quantity throughout the system's lifetime. The evolution of the spin potential (and possible dissipative corrections) is then governed by spin hydrodynamics.

While the equations of spin hydrodynamics can be formulated in several ways~\cite{Fukushima:2020ucl,Florkowski:2017ruc,Florkowski:2018fap,Bhadury:2020puc,Bhadury:2020cop,Cao:2022aku,Weickgenannt:2023nge,Weickgenannt:2024ibf,Li:2020eon,Peng:2021ago,Hattori:2019lfp,Kumar:2018iud,Daher:2022xon,Shi:2020htn,Bhadury:2022ulr,Gallegos:2021bzp,She:2024rnx,Montenegro:2017rbu,Biswas:2023qsw,Drogosz:2024gzv,Tiwari:2024trl,Kiamari:2023fbe,Singh:2022ltu,Kumar:2018lok,Weickgenannt:2022qvh,Garbiso:2020puw,Weickgenannt:2022zxs,She:2021lhe,Wagner:2024fry,Chiarini:2024cuv}, the most popular approach is to derive them from quantum kinetic theory (QKT), which describes a quantum-field theoretical system in the limit where the fields are sufficiently localized to be treated in terms of quasi-particles. Utilizing this framework, one possible approach consists in taking the spin tensor to be conserved~\cite{Florkowski:2017ruc,Florkowski:2018fap,Florkowski:2019qdp,Florkowski:2021wvk}, implying that the energy-momentum tensor is symmetric. A numerical solution of spin hydrodynamics derived in such a way was presented in Ref.~\cite{Singh:2024cub} with the spin tensor taken to be in de Groot, van Leeuwen, and van Weert (GLW) pseudogauge. For a suitable choice of initial spin potential, their results show that using a delayed initial time for spin hydrodynamics can successfully describe polarization measurements. This is suggestive of strong dissipative effects during the initial stages of the fireball that may be responsible for the relaxation of the spin potential toward its equilibrium value given by the thermal vorticity. Indeed, it is also possible to derive these dissipative effects in a quantum kinetic framework. In Ref. \cite{Wagner:2024fhf}, it has been found that the relaxation of the spin potential toward the thermal vorticity is sourced by so-called nonlocal collisions \cite{Weickgenannt:2021cuo,Wagner:2022amr}. Intuitively, this is sensible since, in order to exchange orbital angular momentum and spin, the collision cannot occur at a single space-time point \cite{Florkowski:2018fap}. On a more technical side, these collisions essentially depend on the spin-changing matrix elements of the underlying theory, and the equations solved in Ref. \cite{Singh:2024cub} refer to a system in which these interactions can be neglected. In this work, we go beyond this approximation and numerically solve the equations of dissipative spin hydrodynamics. The hydrodynamic and spin degrees of freedom are evolved simultaneously at each timestep. Since our aim is to apply this framework to heavy-ion collisions, we work in the small polarization limit, where the spin sector does not significantly back-react on the hydrodynamic background, effectively decoupling the two sectors at leading order. The aim of this work is not to provide a full statistical analysis, but rather to demonstrate how successfully and consistently our framework can describe the spin-polarization data under the same hydrodynamic background conditions as those used in the equilibrium approach. This direct comparison allows us to assess the impact of going beyond the isothermal approximation and including dissipative spin dynamics.

{\it Notation:} We work in natural units where $\hbar=k_B=c=1$. We adopt the mostly negative metric convention, $g_{\mu\nu}=\text{diag}(1,-1,-1,-1)$. The inner product of two four-vectors is denoted by $A\cdot B = g_{\mu\nu}A^\mu B^\nu$. The fluid velocity is denoted by $u^\mu$ such that $u\cdot u = 1$. The operator  $\Delta^{\mu\nu}=g^{\mu\nu}-u^\mu u^\nu$  projects a vector onto the subspace orthogonal to $u^\mu$, which we denote by $A^{\langle \mu\rangle}=\Delta^{\mu\nu}A_\nu$. Similarly, $\Delta_{\alpha \beta}^{\mu\nu} = \frac{1}{2}(\Delta_\alpha^\mu \Delta_\beta^\nu+\Delta_\beta^\mu\Delta_\alpha^\nu -\frac{2}{3} \Delta^{\mu\nu} \Delta_{\alpha \beta})$ projects a rank-2 tensor onto the symmetric and traceless subspace orthogonal to the fluid velocity, which we denote by $A^{\langle \mu\nu\rangle}=\Delta^{\mu\nu}_{\alpha\beta}A^{\alpha\beta}$.
Symmetric and antisymmetric parts of a rank-two tensor  $A^{\mu\nu}$ are represented by $A^{(\mu\nu)}=A^{\mu\nu}+A^{\nu\mu}$ and $A^{[\mu\nu]}=A^{\mu\nu}-A^{\nu\mu}$, respectively. The comoving derivative $u\cdot \partial$ of a quantity $A$ is denoted by $\dot{A}=u\cdot \partial A$. The derivative orthogonal to the 4-velocity is written as $\nabla^\mu =\Delta^{\mu\nu}\partial_\nu$. The covariant derivative of a quantity $A$ is denoted by $D_\mu A$. For the Levi-Cività tensor $\epsilon^{\mu\nu\alpha\beta}$, we follow the sign convention $\epsilon^{0123} =-\epsilon_{0123} = +1$. We also define the following gradients of the fluid velocity: The expansion scalar $\theta=D\cdot u$, the shear tensor $\sigma^{\mu\nu}=D^{\langle\mu}u^{\nu\rangle}$, as well as the (kinematic) vorticity tensor $\omega_K^{\mu\nu}=\frac12\Delta^{\mu}_\alpha \Delta^\nu_\beta D^{[\alpha}u^{\beta]}=\epsilon^{\mu\nu\alpha\beta}u_\alpha \omega_{K,\beta}$, with $\omega_K^\mu$ being the associated vorticity vector.
    
\section{Theoretical background}
The degrees of freedom quantifying the spin density of the fluid can be described via the spin potential $\Omega^{\mu\nu}_0$, which constitutes an antisymmetric  rank-two tensor and can thus be decomposed into electric and magnetic parts,\footnote{Note that the sign of $\kappa_0^\mu$ differs between Ref. \cite{Florkowski:2018fap} and this work, which uses the same conventions as Refs. \cite{Wagner:2024fhf, Wagner:2024fry}.}
\begin{equation}
\Omega_0^{\mu\nu}=u^{[\mu}\kappa_0^{\nu]}+\epsilon^{\mu\nu\alpha\beta}u_\alpha\omega_{0,\beta}\;.
\end{equation}
Furthermore, when considering dissipative spin hydrodynamics, it has been shown in Ref. \cite{Wagner:2024fry} that one has to consider an additional rank-2 tensor $\t^{\mu\nu}$ which has the same symmetries as the shear-stress tensor $\pi^{\mu\nu}$ and can thus be referred to as the spin-shear stress. In the GLW pseudogauge, it appears in the spin tensor as $S^{\mu,\alpha\beta}\sim \mathfrak{t}^{\mu[\alpha}u^{\beta]}$.\footnote{Note that the equations of motion for the spin degrees of freedom $\omega_0^\mu$, $\kappa_0^\mu$, and $\mathfrak{t}^{\mu\nu}$ follow from (a truncation of) the moments of the quantum-kinetic Boltzmann equation and are independent of the definition of macroscopic currents such as the spin tensor.}

The evolution equations for the (macroscopic) spin degrees of freedom ($\omega_{0}^\mu,\ \kappa_{0}^\mu,\ \mathfrak{t}^{\mu\nu}$), which do not influence the background hydrodynamics in the limit of small polarization, can be derived from QKT employing the inverse-Reynolds dominance (IReD) approach \cite{Wagner:2022ayd}, and are given by \cite{Wagner:2024fry} (for the case of zero baryon diffusion)
\begin{widetext}
\begin{subequations}
\begin{align}
\tau_\omega \dot{\omega}_0^{\langle\mu\rangle}+\omega_0^\mu &= - \frac{\omega_{\text{K}}^\mu}{T} + \epsilon^{\mu\nu\alpha\beta}u_\nu \left(\ell_{\omega\kappa}\nabla_\alpha \kappa_{0,\beta}
-\tau_{\omega}\dot{u}_\alpha \kappa_{0,\beta}\right)+\delta_{\omega\omega}\omega_0^\mu \theta + \lambda_{\omega\omega}\sigma^{\mu\nu}\omega_{0,\nu}
+\lambda_{\omega \t} \t^{\mu}{}_\nu\omega_{\text{K}}^\nu \label{eq:eom_omega_hydro} \;,\\
\tau_\kappa \dot{\kappa}_0^{\langle\mu\rangle}+\kappa_0^\mu &= -\frac{\dot{u}^\mu}{T} + \epsilon^{\mu\nu\alpha\beta}u_\nu \left(\frac{\tau_\kappa}{2}\nabla_\alpha \omega_{0,\beta}+\tau_{\kappa}\dot{u}_\alpha \omega_{0,\beta}\right)+\delta_{\kappa\kappa}\kappa_0^\mu \theta 
+\left(\lambda_{\kappa\kappa} \sigma^{\mu\nu}
+\frac{\tau_\kappa}{2} \omega_{\text{K}}^{\mu\nu}\right)\kappa_{0,\nu}
 \nonumber\\
&\qquad \quad \;\, +\tau_{\kappa \t}\t^{\mu\nu}\dot{u}_\nu +\ell_{\kappa \t}\Delta^\mu_\lambda \nabla_\nu \t^{\nu\lambda}  \label{eq:eom_kappa_hydro} \;,\\
\tau_\t \dot{\t}^{\langle\mu\nu\rangle}+\t^{\mu\nu} &= \frac{\mathfrak{d}}{T}\sigma^{\mu\nu} + \delta_{\t\t} \t^{\mu\nu} \theta 
+\lambda_{\t\t}\t_\lambda{}^{\langle\mu}\sigma^{\nu\rangle\lambda}
+\frac53\tau_{\t}\t_\lambda{}^{\langle\mu}\omega_{\text{K}}^{\nu\rangle\lambda}
+\ell_{\t \kappa} \nabla^{\langle\mu} \kappa_0^{\nu\rangle}
+\tau_{\t\omega} \omega_{\text{K}}^{\langle\mu}\omega^{\nu\rangle}_0
+\lambda_{\t\omega} \sigma_\lambda{}^{\langle\mu}\epsilon^{\nu\rangle\lambda\alpha\beta} u_\alpha \omega_{0,\beta} \label{eq:eom_t_hydro} \;.
\end{align}
\end{subequations}
\end{widetext}
\begin{figure*}[t]
	\centering
	\includegraphics[scale=0.8]{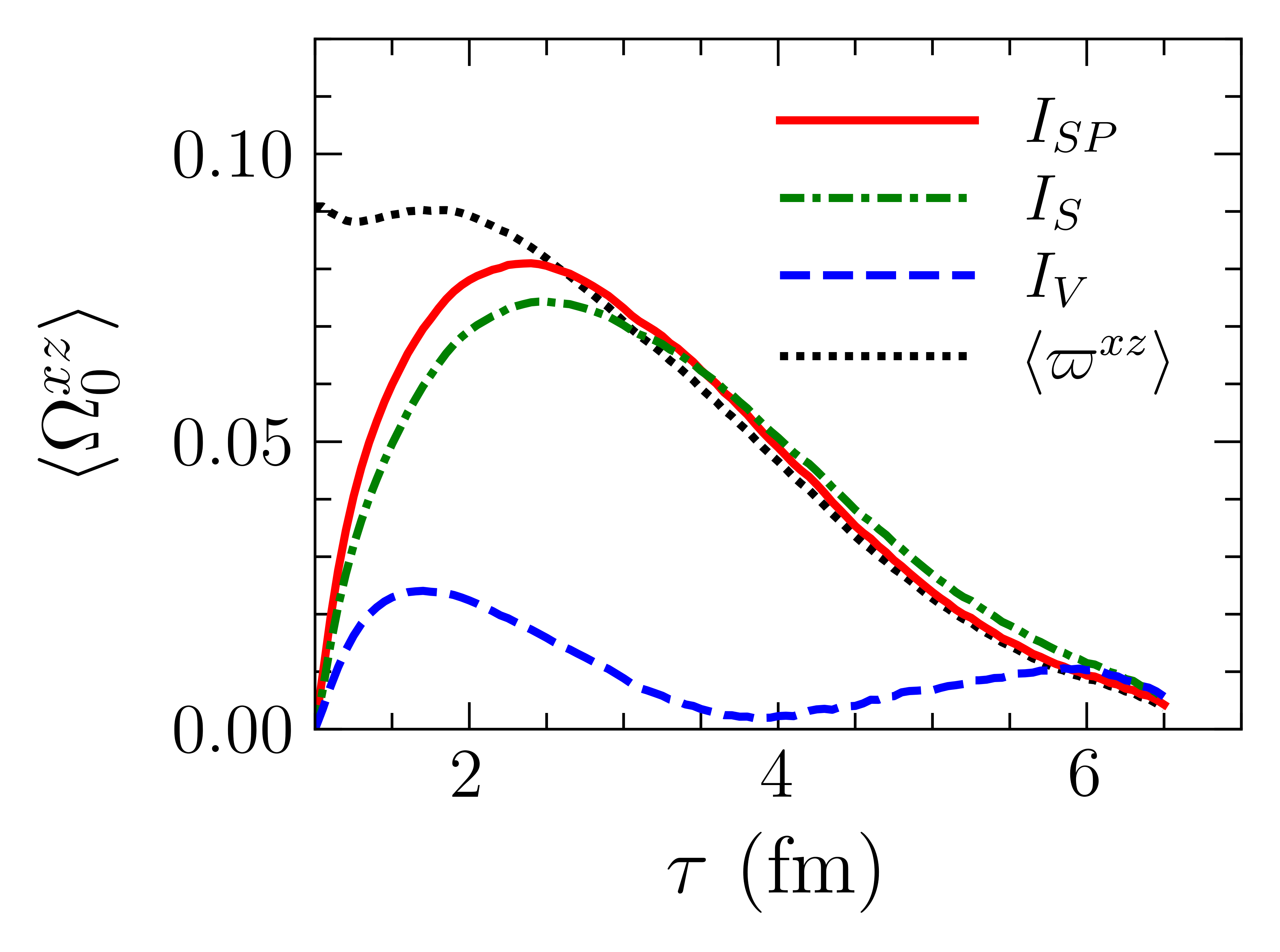}\hspace*{0.1\textwidth}
	\includegraphics[scale=0.8]{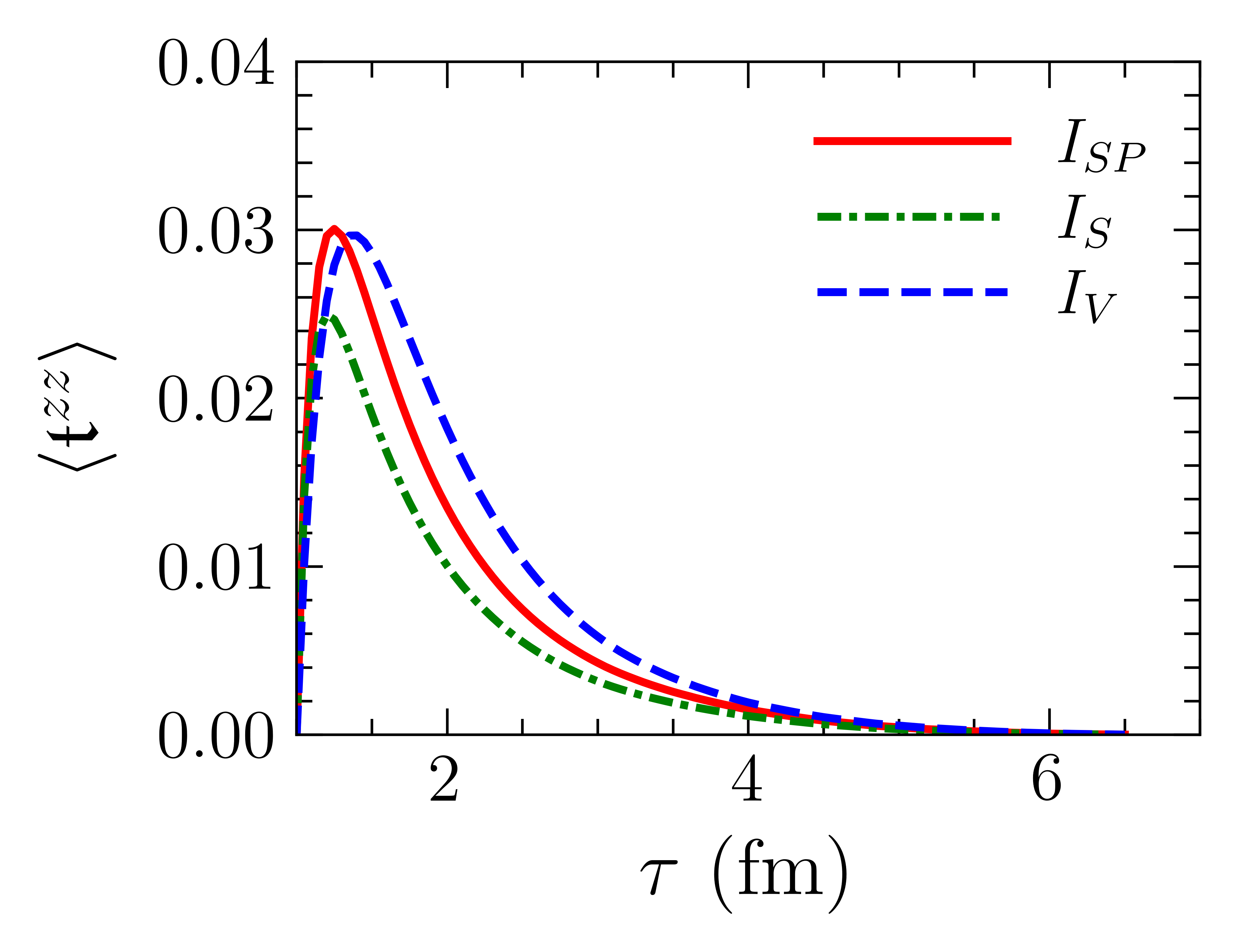}
	\caption{Time evolution of the energy-weighted space-averaged (left) $xz$ component of the spin potential, and (right) $zz$ component of the spin-shear stress, for three different interactions. Due to the symmetry of our initial condition model, the spatial average of all other components of the spin potential is zero.}
	\label{fig:spinpotevln}
\end{figure*}
The relaxation times $\tau_\omega$ and $\tau_\kappa$ quantify the time scales on which $\omega_0^\mu$ and $\kappa_0^\mu$ approach their asymptotic values, i.e., the first terms on the right-hand sides of Eqs. \eqref{eq:eom_omega_hydro} and \eqref{eq:eom_kappa_hydro}. As mentioned in the introduction, these times are (inversely) proportional to the nonlocal collision terms. 
On the other hand, the coefficient $\tau_\mathfrak{t}$, which sets the time scale of $\mathfrak{t}^{\mu\nu}$, is (inversely) proportional to local collisions, indicating that the relaxation of the spin-shear stress is not driven solely through spin-orbit coupling.
The asymptotic value of the spin-shear stress, given by the first term on the right-hand side of Eq.~\eqref{eq:eom_t_hydro}, is proportional to the shear tensor, reminiscent of the thermal-shear contribution in Ref. \cite{Becattini:2021suc}. The corresponding transport coefficient $\mathfrak{d}$ is given by a ratio of nonlocal and local collision terms. In particular, while it depends on the underlying microscopic interaction, it is independent of its absolute strength, and vanishes when the collisions are local.  

\section{Numerical framework}
Solving Eqs.~\eqref{eq:eom_omega_hydro}-\eqref{eq:eom_t_hydro} requires $T$, $u^\mu$, and their gradients from the background hydrodynamics. The details of the background hydrodynamics are discussed in Appendix~\ref{app: hydro}. We use the code developed in Refs.~\cite{Singh:2024cub,Singh2023hdo,Singh2023qcp} to solve the equations of background hydrodynamics numerically in (3+1)-dimensions in Milne coordinates $(\tau,x,y,\eta_s)$, where $\tau=\sqrt{t^2-z^2}$ and $\eta_s=\frac12\ln \left(\frac{t+z}{t-z}\right)$. The code, developed in \texttt{Forran}, incorporates the Godunov-type relativistic Harten-Lax-van Leer-Einfeldt (HLLE) approximate Riemann solver for the computation of numerical fluxes at fluid cell boundaries, as described in Refs.~\cite{RISCHKE1995346,Karpenko:2013wva}. The implementation of Eqs.~\eqref{eq:eom_omega_hydro}-\eqref{eq:eom_t_hydro} is inspired by the publicly available \texttt{vHLLE} code~\cite{Karpenko:2013wva,vhlleweblink}. Additionally, the \texttt{FORTRAN} code supports parallel programming on multiple threads using \texttt{OpenMP}~\cite{openmpweblink}, a feature not present in the public version of \texttt{vHLLE}. The equations of background hydrodynamics are closed with an equation of state for which we use the lattice-QCD-based equation of state NEOS-BQS~\cite{Monnai2019eos,HotQCD:2014kol,HotQCD:2012fhj,Ding:2015fca,Bazavov:2017dus,eosweblink}. We extend the code to solve the spin equations, Eqs.~\eqref{eq:eom_omega_hydro}-\eqref{eq:eom_t_hydro}, with the numerical details provided in Appendix~\ref{sec:numerics}.

For the computation of the transport coefficients in Eqs.~\eqref{eq:eom_omega_hydro}-\eqref{eq:eom_t_hydro}, one has to provide an explicit model of the system. Deferring the use of a kinetic framework with all particle species included to future work, we employ a simplified treatment by assuming a system of spin-\nicefrac{1}{2} particles with a mass of 300 MeV, which is on the order of the thermal masses $m_\text{th}\sim gT$ of the particles in the hot and dense fireball. We consider three types of interactions among the particles, namely (i) scalar, where the interaction Lagrangian is $\mathcal{L}_\text{int,S}= G(\bar{\psi}\psi)^2$, (ii) scalar+pseudoscalar, where $\mathcal{L}_\text{int,SP}=G[(\bar{\psi}\psi)^2-(\bar{\psi}\gamma_5\psi)^2]$,\footnote{This interaction is equivalent to the well-known Nambu-Jona--Lasinio (NJL) model~\cite{Nambu:1961tp}.} and (iii) vector, with $\mathcal{L}_\text{int,V}=-G(\bar{\psi}\gamma^\mu\psi)^2$, with a coupling constant $G$. We denote these three scenarios by $I_S$, $I_{SP}$, and $I_V$, respectively. As mentioned earlier, the transport coefficient $\mathfrak{d}$ depends only on the form of the interaction but is independent of the coupling constant $G$. The same can be said for the relaxation times $\tau_\omega$, $\tau_\kappa$, and $\tau_\t$, if we express them in units of the shear-stress relaxation time $\tau_\pi$. Consequently, we tabulate the relaxation times $\tauw$, $\tauk$, and $\taut$ as functions of $z=m/T$ in units of $\tau_\pi$ for the three types of interactions mentioned above (see Fig.~\ref{fig:spintranscoeff}). All other transport coefficients in the spin equations are tabulated as functions of $z$ in units of the corresponding spin relaxation times and isotropic pressure to make them dimensionless (see Ref.~\cite{Wagner:2024fry} for details). We assume that these dimensionless ratios are universal, depending only on the nature of the interaction between particles, and thus remain unchanged for the quark-gluon fluid formed in relativistic HIC. Consequently, for transport coefficients that involve pressure, we use the local thermodynamic equilibrium pressure, $P$, of the fireball to extract their numerical values from tables of dimensionless ratios. This approach aligns with standard hydrodynamic simulations, such as MUSIC~\cite{musicweblink}, where second-order transport coefficients computed using kinetic theory in the massless limit are assumed to apply to the quark-gluon fluid. To solve Eqs.~\eqref{eq:eom_omega_hydro}-\eqref{eq:eom_t_hydro}, we initialize $\omega_{0}^\mu$, $\kappa_{0}^\mu$, and $\mathfrak{t}^{\mu\nu}$ to zero. This choice of initial condition for the spin degrees of freedom assumes that the total angular momentum at the beginning of the hydrodynamic evolution of the fireball is dominated solely by the orbital component.

The formula for the spin vector can be written as
\begin{equation}
\label{eq:spinvec}
  S^\mu (p) = S^\mu_\omega (p) + S^\mu_\kappa (p) + S^\mu_{\mathfrak{t}} (p)\;,
\end{equation}
where~\cite{Wagner:2024fry}
\begin{subequations}
\begin{align}
  S^\mu_\omega (p) &= \frac{1}{N(p)}\int \d\Sigma\cdot p \frac{u^\mu (\omega_0\cdot p) -\omega_0^\mu(p\cdot u)}{2 m_\Lambda} f_0\widetilde{f}_0\label{eq:polw},\\
  S^\mu_\kappa (p) &= -\frac{1}{N(p)}\int \d\Sigma\cdot p  \frac{\epsilon^{\mu\nu\rho\sigma}u_\nu p_\sigma }{2m_\Lambda}\kappa_{0,\rho} f_0\widetilde{f}_0\label{eq:polk}\;,\\
  S^\mu_{\mathfrak{t}} (p) &= \frac{1}{N(p)}\int \d\Sigma\cdot p \frac{\epsilon^{\mu\nu\rho\sigma}u_\nu p^\lambda p_\sigma}{3T^2(\varepsilon+P)} \mathfrak{t}_{\rho\lambda} f_0\widetilde{f}_0\label{eq:polt}\;,
\end{align}
\end{subequations}
where $m_\Lambda = 1.115$ GeV, $f_0$ denotes the Fermi-Dirac distribution, $\widetilde{f}_0=1-f_0$, and
$N(p)=2\int \d\Sigma\cdot p\,  f_0$. We carry out the spin polarization analysis on the particlization (or switching) hypersurface defined by $\varepsilon_{\rm sw}$=0.5 GeV/fm$^3$.

\begin{figure*}[t]
\centering
\includegraphics[scale=0.8]{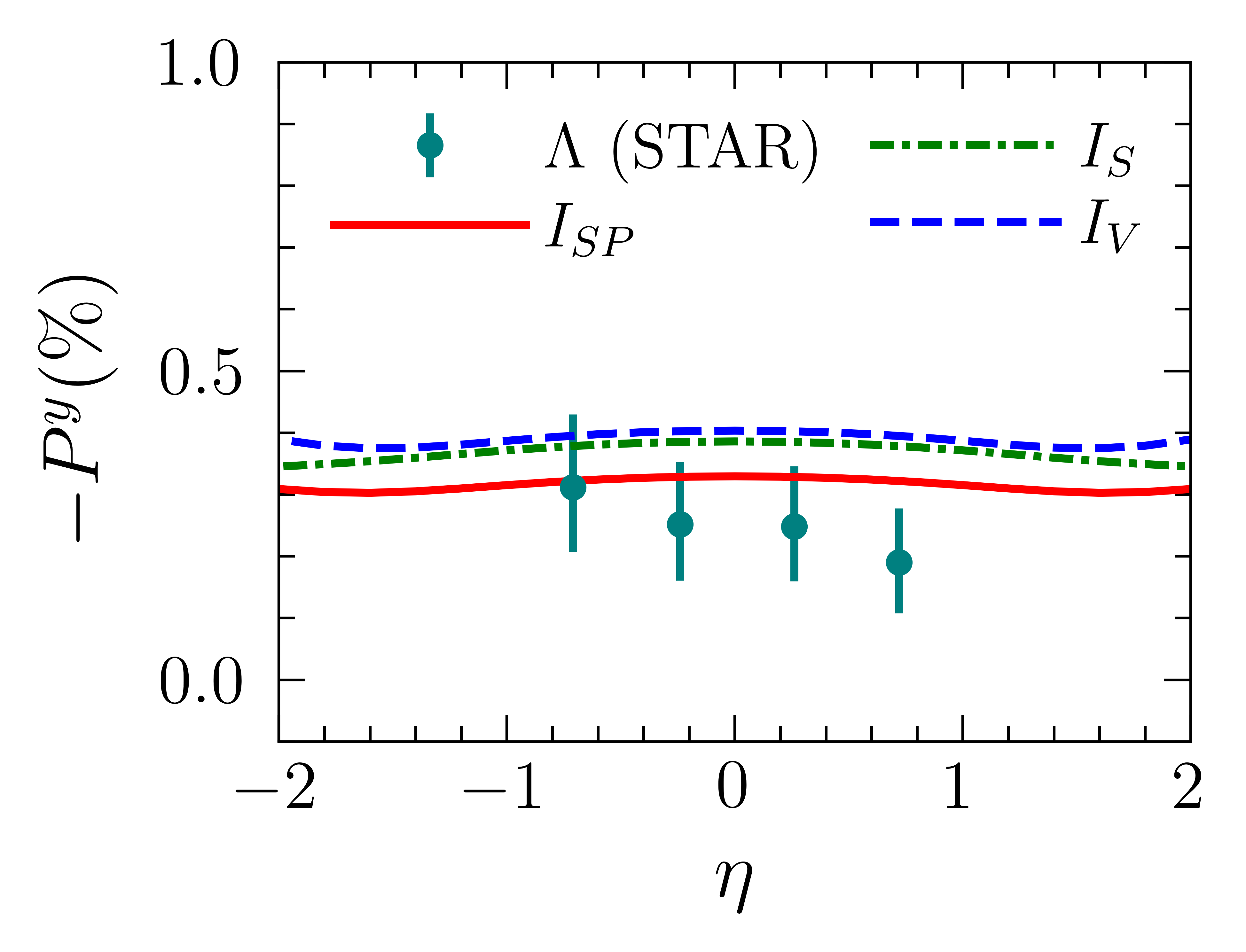}\hspace*{0.6in}
\includegraphics[scale=0.8]{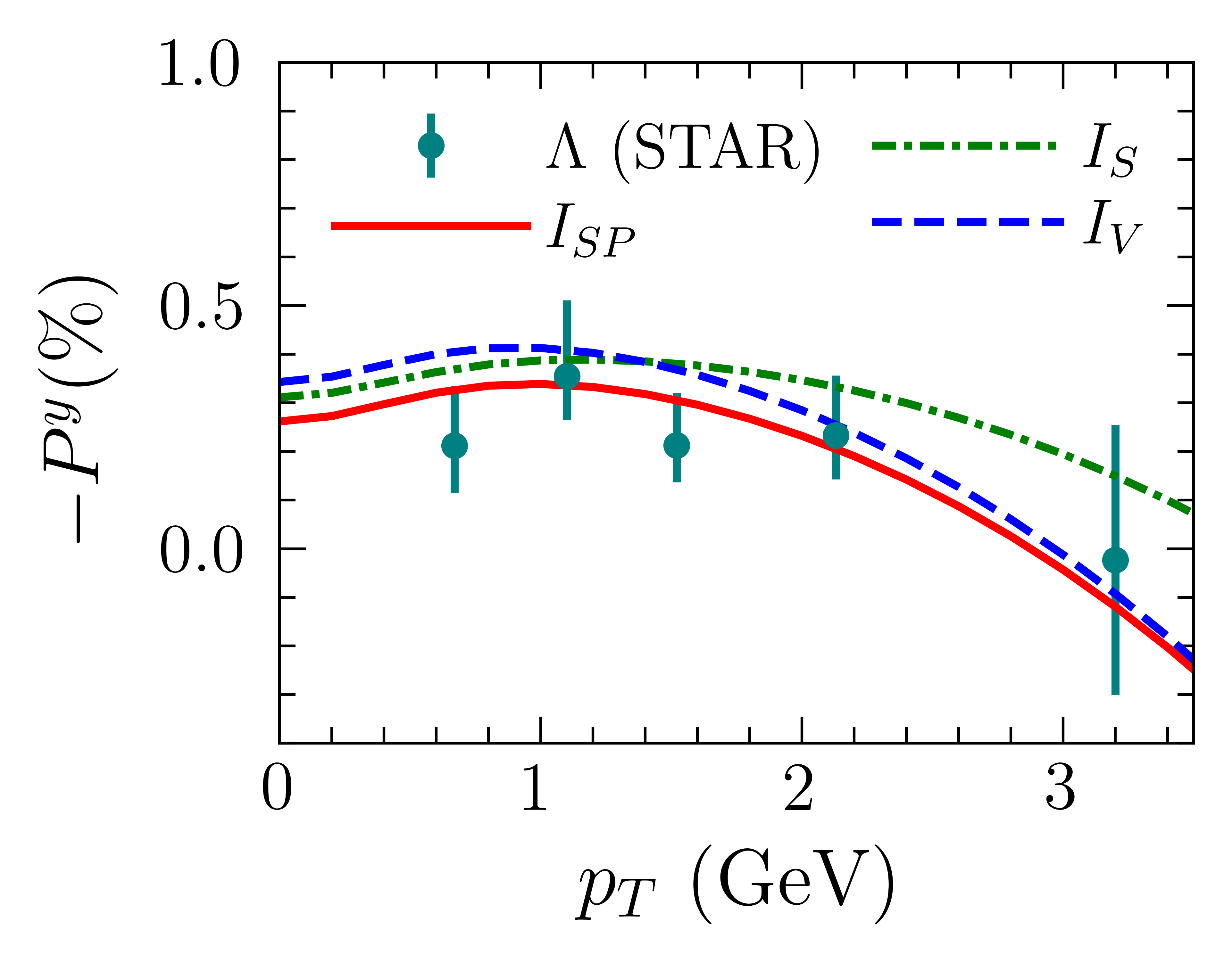}
\includegraphics[scale=0.8]{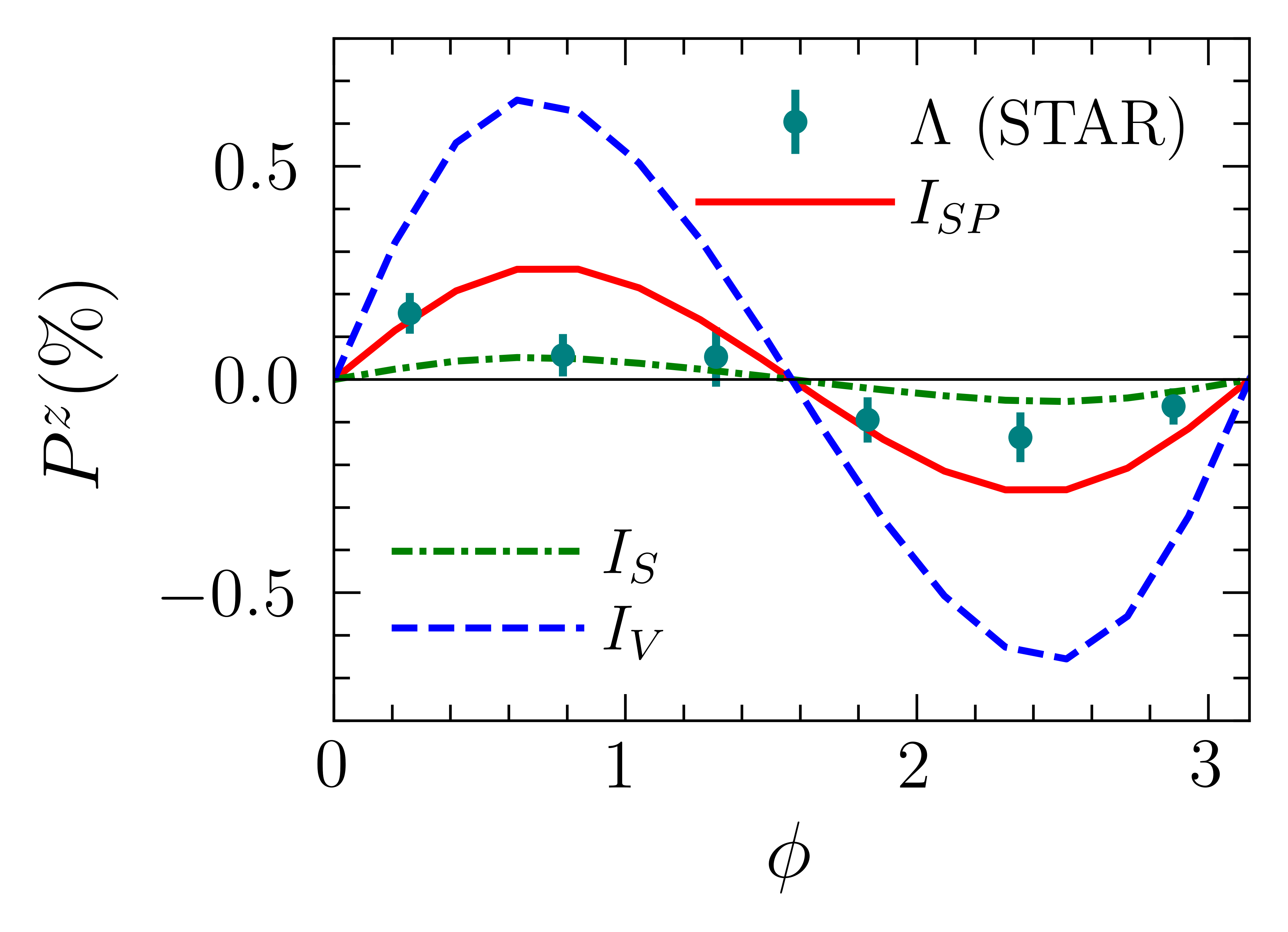}\hspace*{0.55in}
\includegraphics[scale=0.8]{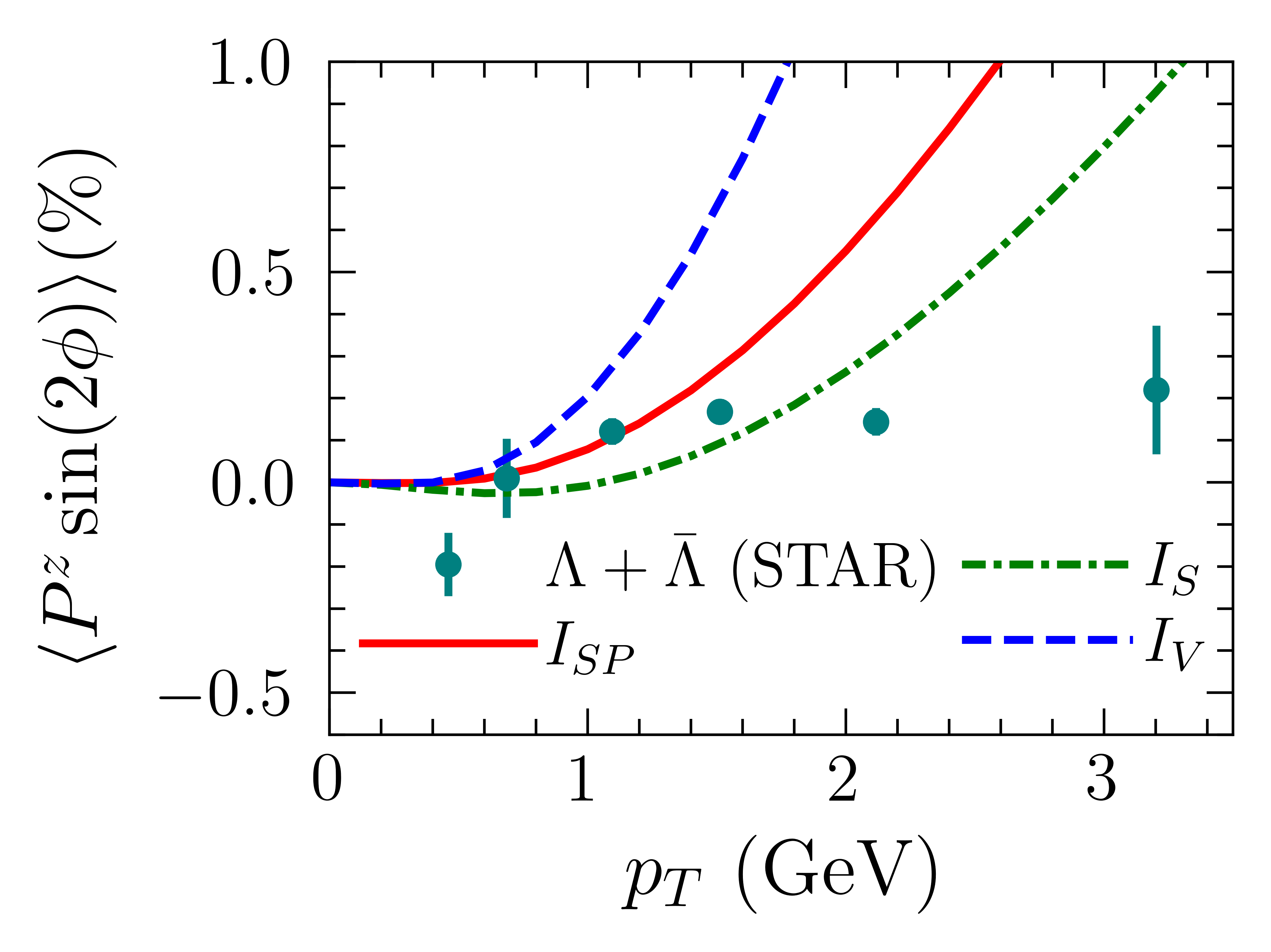}
\caption{Comparison of results from dissipative spin hydrodynamics with experimental data for three interaction scenarios for Au+Au collisions at $\sqrt{s_{NN}}=200$ GeV. The simulations are performed with fixed impact parameter, $b=8.4$ fm, corresponding to 20-50\% centrality. The top panel shows the (left) pseudorapidity and (right) transverse momentum dependence of the $y$ component of $\Lambda$ polarization. The bottom panel shows the (left) azimuthal and (right) transverse momentum dependence of the $z$ component of $\Lambda$ polarization. Experimental data are taken from Refs.~\cite{STAR200_Pj,STAR:2019erd} with the updated decay parameter ($\alpha_\Lambda = 0.732$).}
\label{fig:spinhydropred}
\end{figure*}

A subtle issue arises concerning the tensor $\mathfrak{t}^{\mu\nu}$ in Eq.~\eqref{eq:polt}. At the particlization surface, where fluid cells convert into hadrons such as $\Lambda$, one has to relate the tensor $\mathfrak{t}^{(\Lambda),\mu\nu}$ pertaining to $\Lambda$ to the tensor $\mathfrak{t}^{\mu\nu}$ describing the fluid. While more sophisticated approaches exist \cite{Molnar:2014fva}, for simplicity, we adopt the ``democratic" Grad Ansatz~\cite{Molnar_2011} that essentially gives all particle species emerging from the fluid an equal share of the total quantity $\mathfrak{t}^{\mu\nu}$.

The spin vector in the rest frame of $\Lambda$, $S^{*\mu} = (0,\textbf{S}^*)$ is obtained from $S^\mu = (S^0,\textbf{S})$ as follows:
\begin{equation*}
    \textbf{S}^* = \textbf{S}-\frac{\textbf{p}.\textbf{S}}{E(E+m_\Lambda)}\textbf{p}\;.
\end{equation*}
Here, $E=p^0$ is the $\Lambda$ energy in the laboratory frame. Finally, the spin polarization of $\Lambda$ is computed as $\textbf{P}=2\textbf{S}^*$.

\section{Results}
With the framework described above, we analyze the behavior of dissipative spin hydrodynamics in HICs. 
In Fig.~\ref{fig:spinpotevln}, we show the time evolution of the energy-weighted, space-averaged spin potential and spin-shear stress (where the averaging is performed only over fluid cells with energy density greater than $\varepsilon_{\rm sw}$) for three different interactions in the fireball of Au+Au collisions at 200 GeV, with an impact parameter of $b = 8.4$ fm corresponding to 20-50\% centrality~\cite{Singh:2024cub}. At late times, the relaxation of the spin potential toward thermal vorticity is governed by the longest timescale, which is $\tauw$ for all three interactions. At early times, the evolution is primarily governed by the transport coefficients $\tau_{\kappa}$, $\mathfrak{d}$, $\tau_{\t}$ and $\ell_{\kappa \t}$. For $I_S$ and $I_{SP}$ cases, smaller values of $\tau_{\kappa}$ make the relaxation term dominant, whereas the peak value of the rise is controlled by $\mathfrak{d}$ and $\tau_{\t}$. However, for the $I_V$ case, we observe significant overdamping at early times due to relatively large values of $\tau_{\kappa}$, $\tau_{\t}$, and $\mathfrak{d}$ compared with the other two interactions. To rule out any numerical artifacts, we performed a series of numerical experiments by tuning individual transport coefficients, and concluded that this is indeed a physical effect. The relatively large value of $\tau_{\kappa}$ reduces the relaxation term's contribution compared with the term involving the coefficient $\ell_{\kappa \t}$, which has a damping effect. This damping persists until the spin-shear stress $\mathfrak{t}^{\mu\nu}$ nearly vanishes at later times, allowing the relaxation term to dominate and drive the system by increasing toward the thermal vorticity.

With a qualitative understanding of spin evolution, we now apply our framework to compute the spin polarization of $\Lambda$ hyperons. We first examine the individual contributions of $\omega^\mu_0$, $\kappa^\mu_0$, and $\mathfrak{t}^{\mu\nu}$ to $\Lambda$ polarization (see Fig.~\ref{fig:spinindividual}) for the three different interactions. We find that the transverse momentum dependence of the $y$ component of polarization exhibits marginal sensitivity to the interaction type, with the spin-shear stress contribution being small at low $p_T$ but increasing at higher $p_T$. In contrast, for the azimuthal angle dependence of the $z$ component, the contribution from $\omega_0^\mu$ is small, and that from $\kappa_0^\mu$ is largely independent of the interaction type. Notably, the spin-shear stress plays a significant role, exhibiting strong sensitivity to the interaction type and behaving similar to the thermal shear in Ref.~\cite{Becattini:2021iol}.

The net local $\Lambda$ polarization, obtained by summing the individual contributions from $\omega^\mu_0$, $\kappa^\mu_0$, and $\mathfrak{t}^{\mu\nu}$, for the interactions considered in this work, is shown in Fig.~\ref{fig:spinhydropred}. The results are compared with experimental data (teal dots). There is rather good agreement with the experimental data on the pseudorapidity ($\eta$) and the transverse momentum ($p_T$) dependence of the $y$ component of the polarization for the $I_{SP}$ interaction. It is to be noted that our current implementation includes only (quasi) particles and not anti-(quasi) particles. As a result, a discrepancy arises between our results and the experimental data on $p_T$ dependence of $P_z$, which corresponds to a combined average of $\Lambda + \bar{\Lambda}$. Our approach also correctly reproduces the sign of the longitudinal polarization for all three interactions. Regarding the utility of spin hydrodynamics in the context of HIC, these results are very encouraging. Our work considers a fully dynamical system initialized with zero spin current, which is subsequently generated and transported through the exchange between orbital and spin angular momentum. In the course of this process, we obtain the correct sign of longitudinal polarization, with the tensor $\mathfrak{t}^{\mu\nu}$ in spin hydrodynamics playing a role analogous to the thermal shear ($\xi^{\mu\nu}$) in local-equilibrium scenarios. For the collision energy considered in this work ($\sqrt{s_{NN}}=200$ GeV), we conclude that the assumption of the spin thermalizing to its equilibrium value (given through the thermal vorticity) by the end of the fireball's lifetime is a reasonable approximation, in agreement with the conclusion of Ref. \cite{Wagner:2024fhf}. However, at lower energies, this may no longer be the case. Crucially, the framework of spin hydrodynamics treated in this work should be applicable to these energies as well, since it captures the non-equilibrium spin dynamics.

\section{Summary}
In this work, we solved the equations of relativistic dissipative spin hydrodynamics on a (3+1)-dimensional relativistic hydrodynamic background. The evolution equations for spin-\nicefrac{1}{2} particles, derived in Ref.~\cite{Wagner:2024fry}, using quantum kinetic theory with nonlocal collisions, were employed in our study. We assumed zero baryon diffusion in both the hydrodynamic background and the spin-evolution equations, and considered the particles to have a constant mass of 300 MeV. While the first of these assumptions can be lifted by including baryon diffusion, weakening the second one requires, e.g., considering temperature-dependent masses, which would introduce additional source terms in the spin evolution equations~\cite{Bhadury:2023vjx}. We leave the study of such effects to future work.

Our results show that the relevant spin degrees of freedom ($\omega_{0}^\mu$, $\kappa_{0}^\mu$ and $\mathfrak{t}^{\mu\nu}$), initially set to zero, are generated through the exchange of orbital and spin angular momentum. The initial orbital angular momentum originates from the background hydrodynamic initial condition, and the spin degrees of freedom subsequently evolve via spin-orbit interactions among particles. The nature of the microscopic interaction is encoded in the transport coefficients of spin hydrodynamics. We have considered three types of interactions between the particles, which primarily affect the relaxation times $\tauw$, $\tauk$, $\taut$, as well as the transport coefficient $\mathfrak{d}$. All other transport coefficients remain independent of interactions when expressed in units of appropriate combinations of the respective relaxation times and pressure. We apply this framework to study spin dynamics in the hot and dense fireball formed in relativistic HICs. For $\sqrt{s_{NN}}=200$ GeV, our results indicate that the spin potential relaxes to the thermal vorticity during the fireball's lifetime for the interactions considered in this work. A nonzero spin-shear stress ($\mathfrak{t}^{\mu\nu}$) accelerates this relaxation while also reducing the peak value of the spin potential, acting as a dissipative effect. The framework is then used to compute the spin polarization of $\Lambda$ hyperons. Our predictions show reasonable agreement with experimental data for both global and local polarization for the $I_{SP}$ interaction scenario. Notably, our approach not only reproduces the correct sign of the longitudinal polarization but also demonstrates its sensitivity to microscopic details, while the global polarization remains only slightly affected by them. The only free parameter in the equations of spin hydrodynamics is the particle mass, highlighting the predictive power of the theory. Our analysis suggests that the inclusion of low temperature fluid cells delays spin relaxation. This effect becomes particularly significant at lower collision energies, where a baryon-rich fireball is produced at low temperatures, leading to longer spin-relaxation times. Consequently, the spin may not fully thermalize within the fireball's lifetime, making our framework especially relevant to such conditions. 

\textit{Acknowledgements}-- The authors thank F. Becattini, E. Grossi, and M. Shokri for fruitful discussions. S.K.S is supported by ICSC - \emph{Centro Nazionale di Ricerca in High Performance Computing, Big Data and Quantum Computing}, funded by the European Union - NextGenerationEU. D.W acknowledges support by the project PRIN2022 Advanced Probes of the Quark Gluon Plasma funded by
``Ministero dell'Università e della Ricerca.'' D.W is
supported by the Deutsche Forschungsgemeinschaft (DFG,
German Research Foundation) through the CRC-TR 211
``Strong-interaction matter under extreme conditions''--
Project No. 315477589–TRR 211, and by the State of
Hesse within the Research Cluster ELEMENTS (Project
No. 500/10.006).

\appendix

\section{Background Hydrodynamics}\label{app: hydro}
The conservation of energy-momentum and charge are expressed through the following continuity equations:
\begin{align}
D_\mu T^{\mu\nu} (x) &= 0\;,
\label{eq:T_cons}\\
D_\mu N^\mu (x) &= 0\;,
\label{eq:N_cons}
\end{align}
where $T^{\mu\nu}$ is the energy-momentum tensor, and $N^{\mu}$ is the charge current. We choose to work in the Landau frame, defined by
\begin{equation}
\label{eq:Lframe}
T^{\mu}{}_{\nu}  u^\nu= \varepsilon\, u^\mu\;,
\end{equation}
where the energy-momentum tensor and the net baryon current are written as
\begin{align}
\label{eq:T_curr} 
T^{\mu\nu} &= \varepsilon u^\mu u^\nu - (P+\Pi) \Delta^{\mu\nu} + \pi^{\mu\nu}\;,\\
\label{eq:N_curr} 
N^{\mu} &=  n_B\,u^\mu + V^\mu\;.
\end{align}

\begin{figure}[t]
	\includegraphics[scale=0.8]{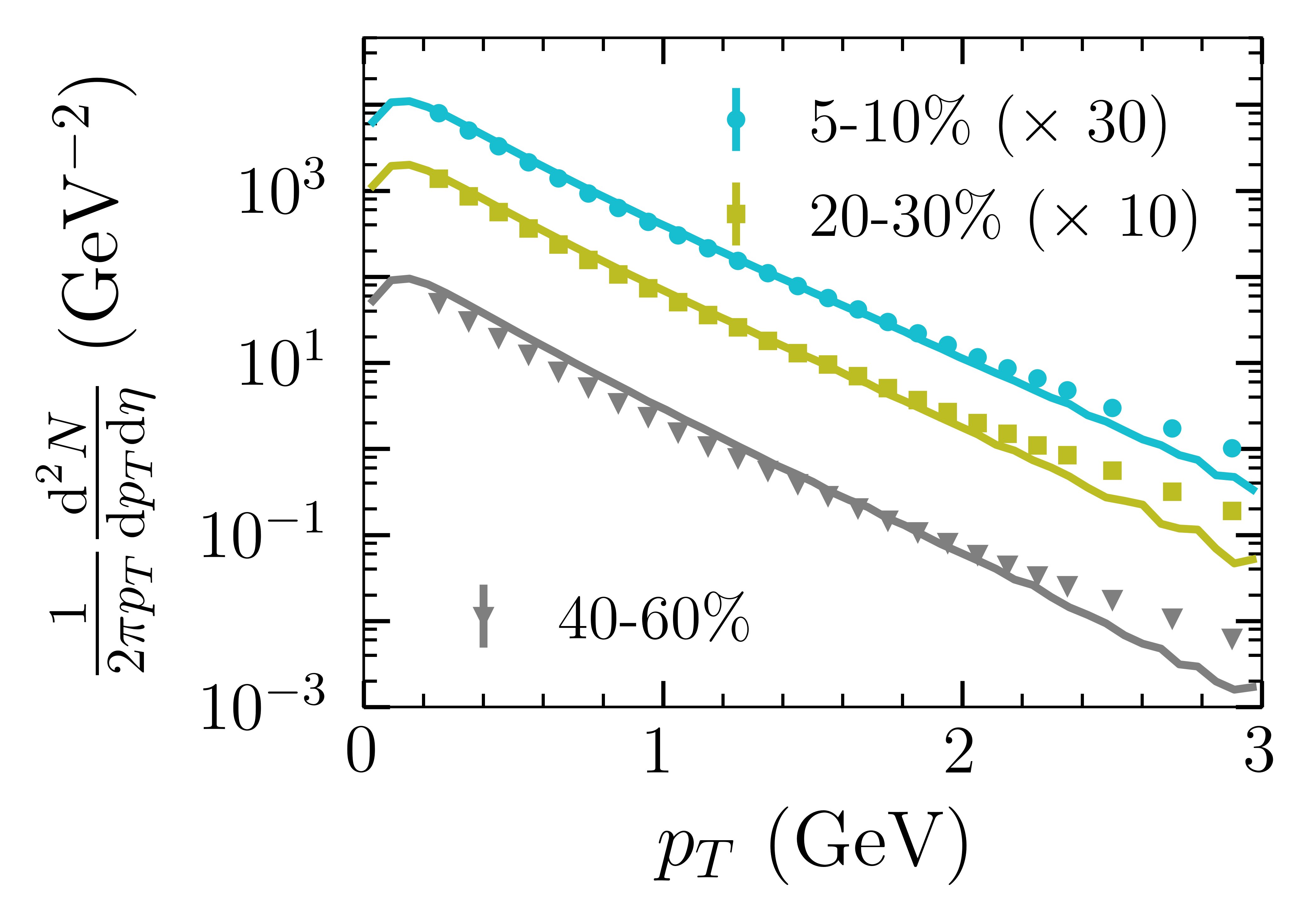}
	\includegraphics[scale=0.8]{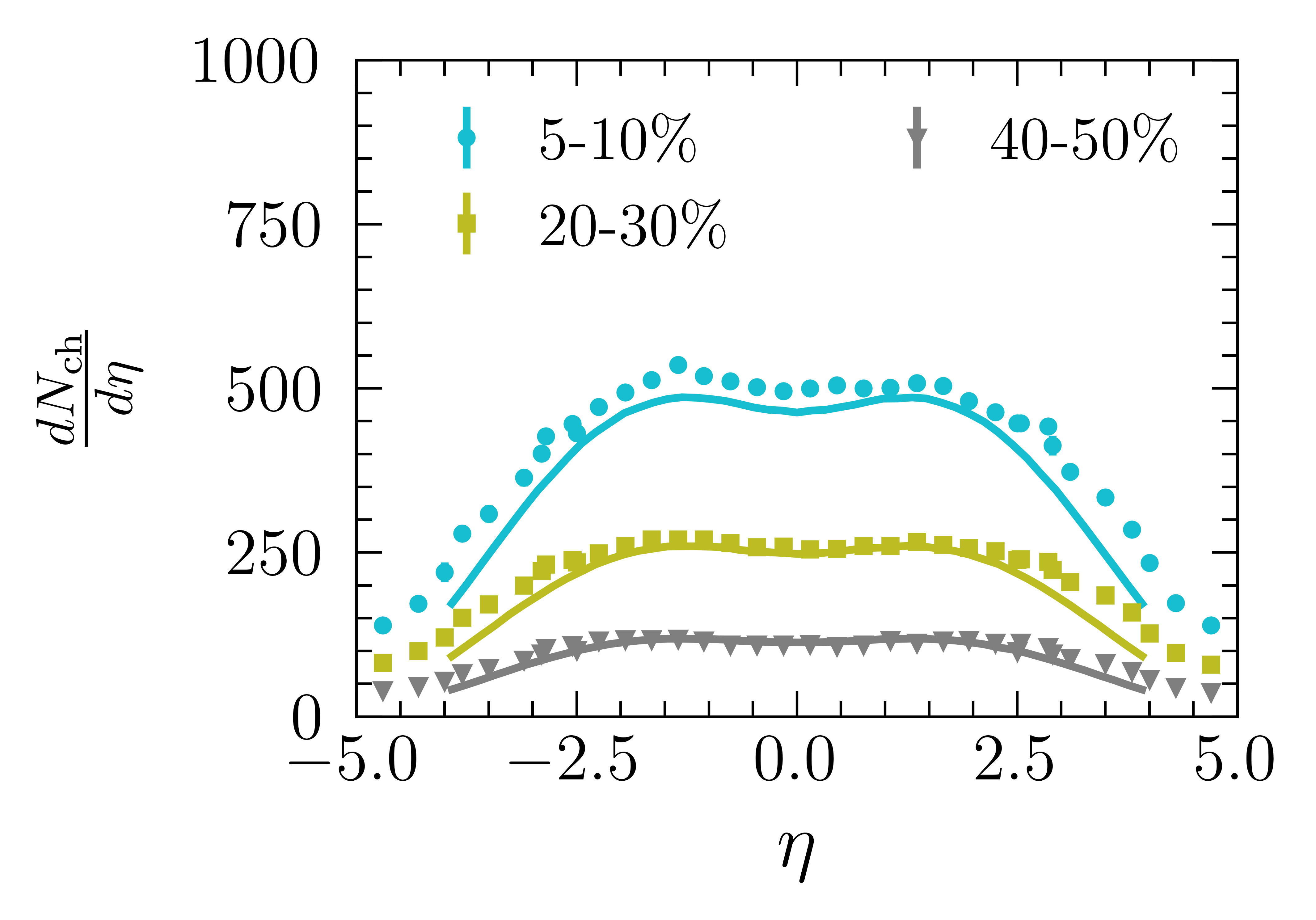}
	\includegraphics[scale=0.8]{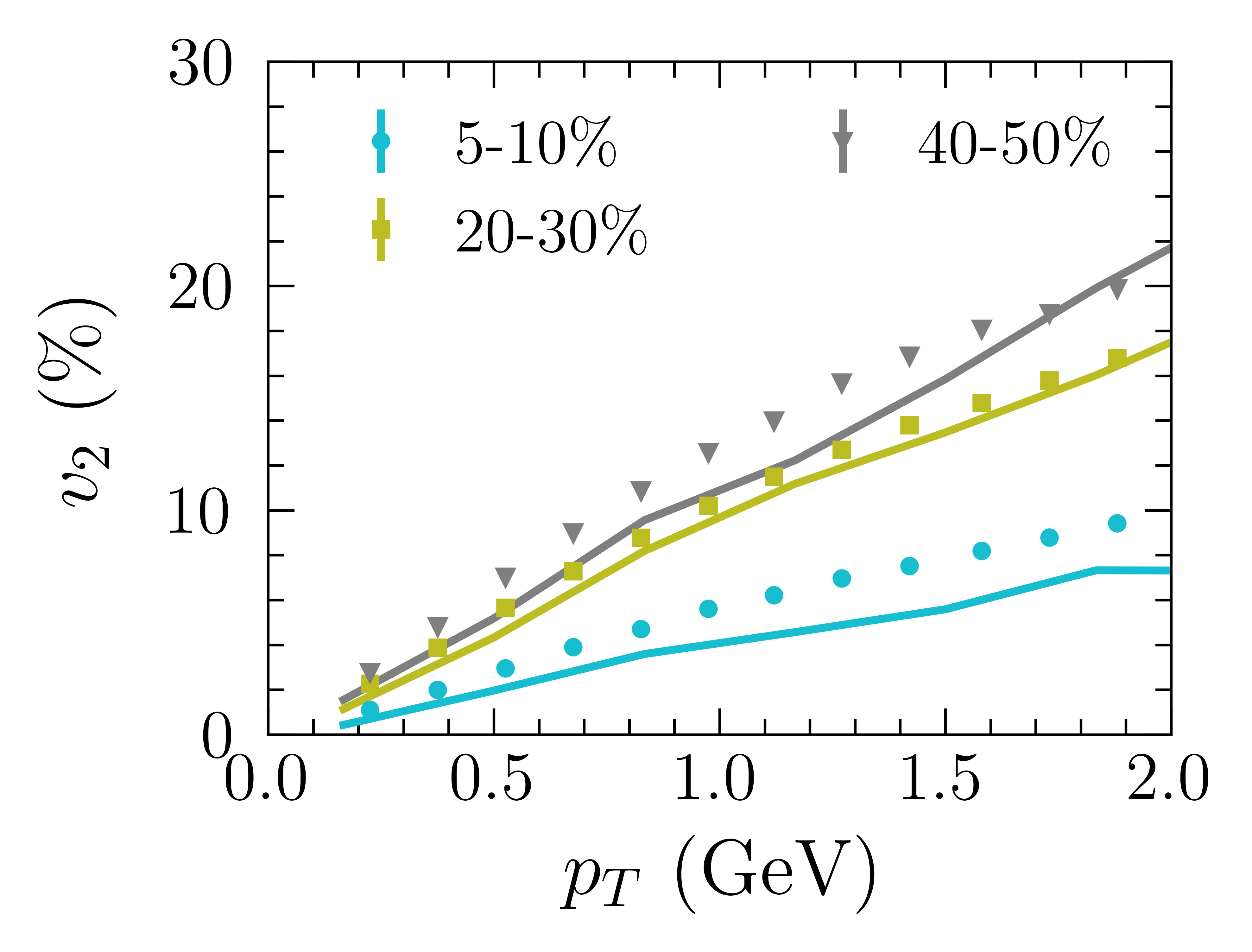}
	\caption{Bulk observables for Au+Au collisions at $\sqrt{s_{NN}}=200$ GeV. Different symbols correspond to experimental data taken from Refs.~\cite{STAR2003_pT,BRAHMS2002_dNdeta,STAR200_v2}.}
	\label{fig:bulk}
\end{figure}

Here $\varepsilon$ is the energy density, $n_B$ is the net-baryon number density, $P$ is the thermodynamic equilibrium pressure, $\pi^{\mu\nu}$ is the shear-stress tensor, $\Pi$ is the bulk viscous pressure, and $V^\mu$ is the baryon diffusion current. In this study, we consider $V^\mu=0$ for simplicity. At first order in velocity gradients, $\pi^{\mu\nu}$ and $\Pi$ are given by their Navier-Stokes values,
\begin{equation*}
\Pi_{\rm NS} = - \zeta D_\gamma u^\gamma = - \zeta \theta \; , \quad \pi_{\rm NS}^{\alpha \beta}= 2\eta\,  \Delta_{\gamma \delta}^{\alpha \beta} D^\gamma u^\delta = 2 \eta\sigma^{\alpha \beta}\;,
\end{equation*}
\begin{figure*}[t]
	\centering
	\begin{minipage}{0.48\linewidth}
		\centering
		\includegraphics[scale=0.8]{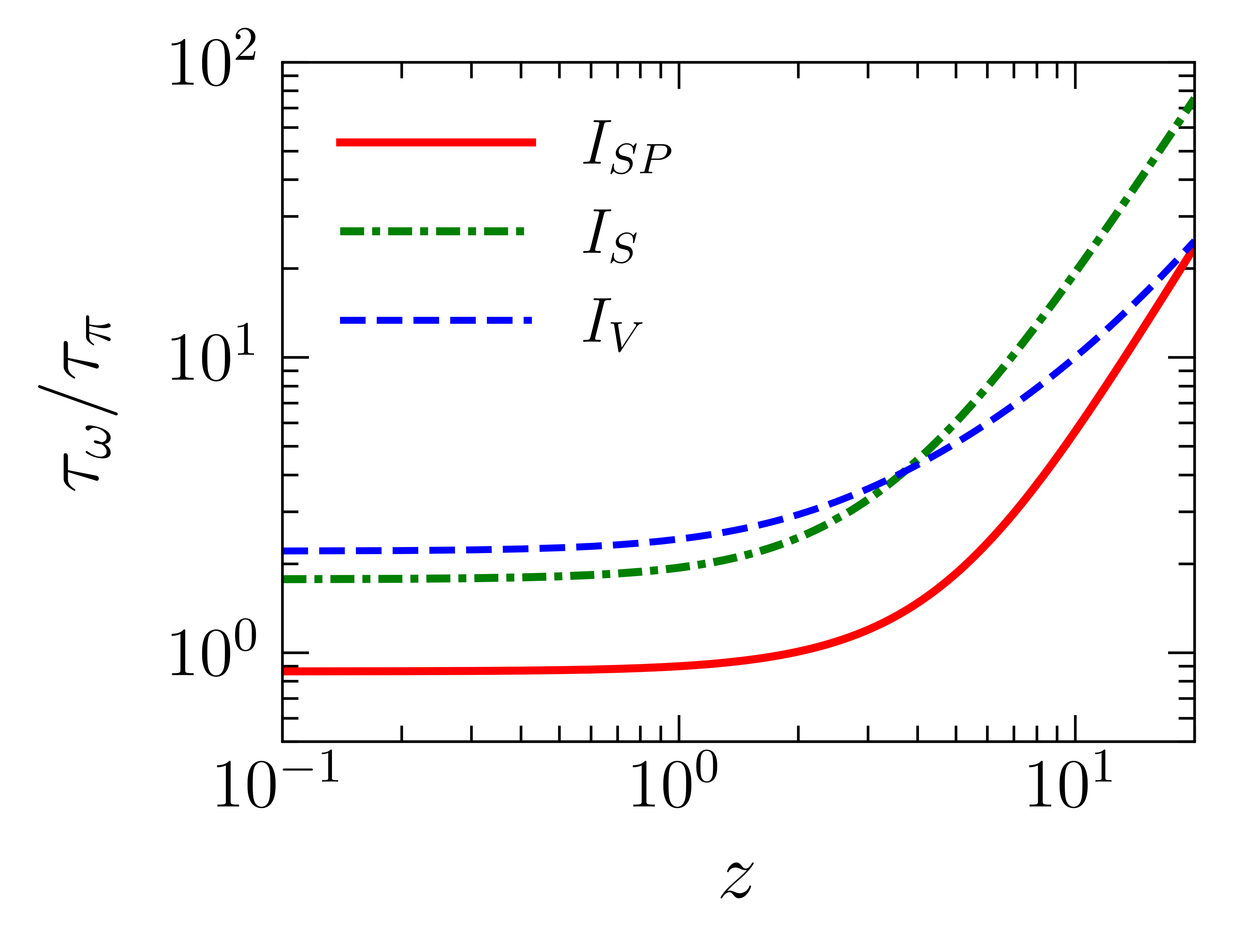}
	\end{minipage}
	\hfill
	\begin{minipage}{0.48\linewidth}
		\centering
		\includegraphics[scale=0.8]{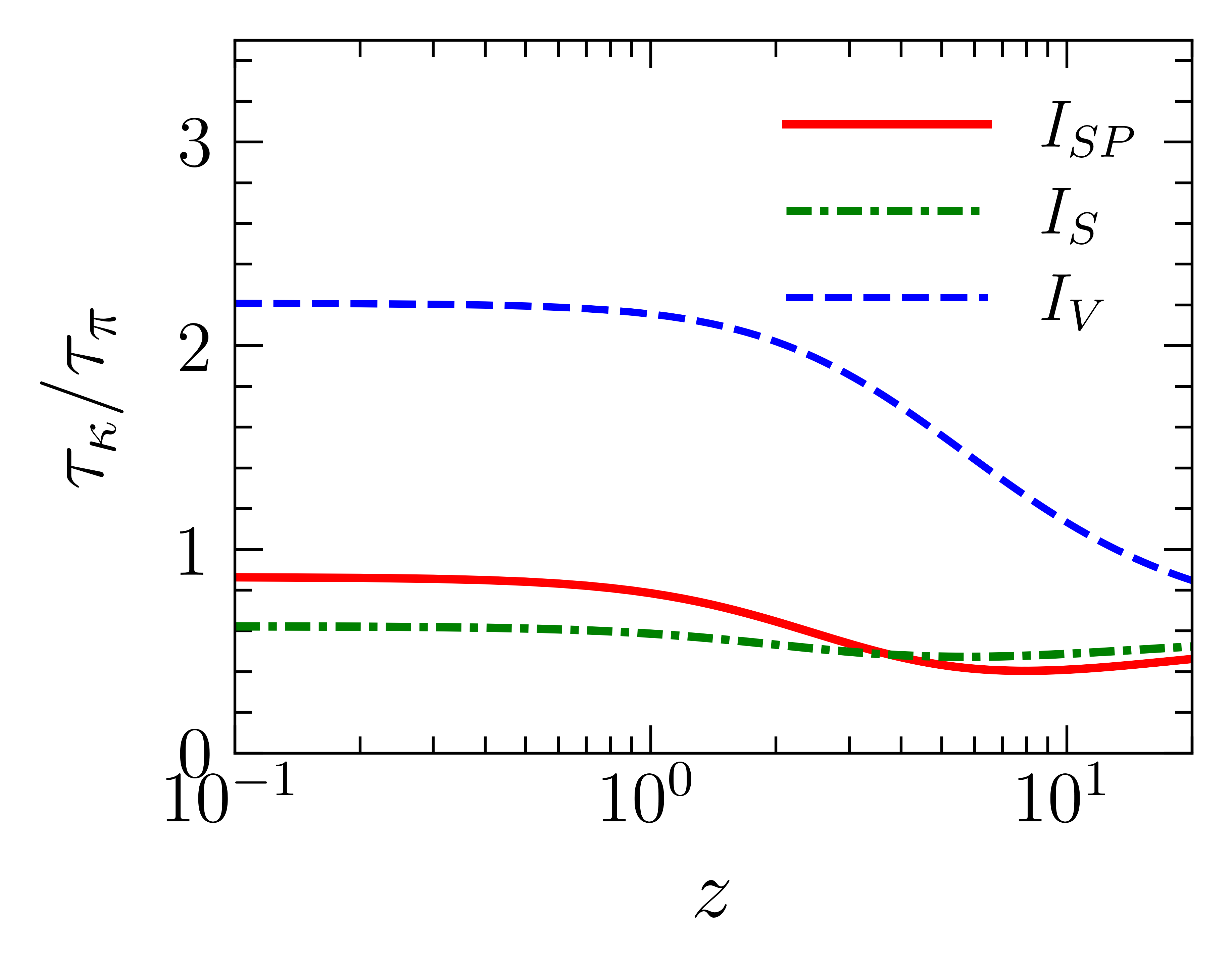}
	\end{minipage}
	\\[1ex] 
	\begin{minipage}{0.48\linewidth}
		\centering
		\includegraphics[scale=0.8]{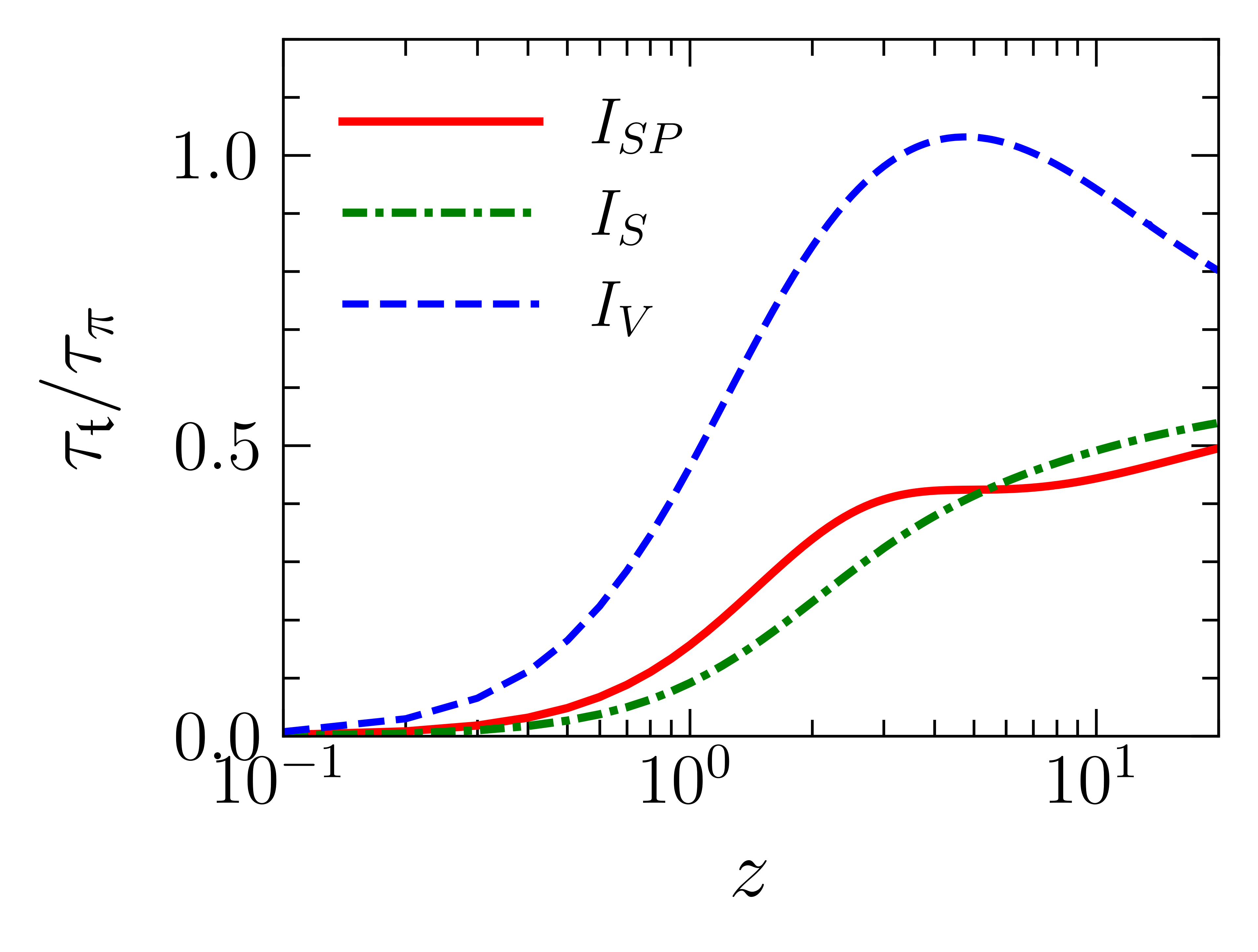}
	\end{minipage}
	\hfill
	\begin{minipage}{0.48\linewidth}
		\centering
		\includegraphics[scale=0.8]{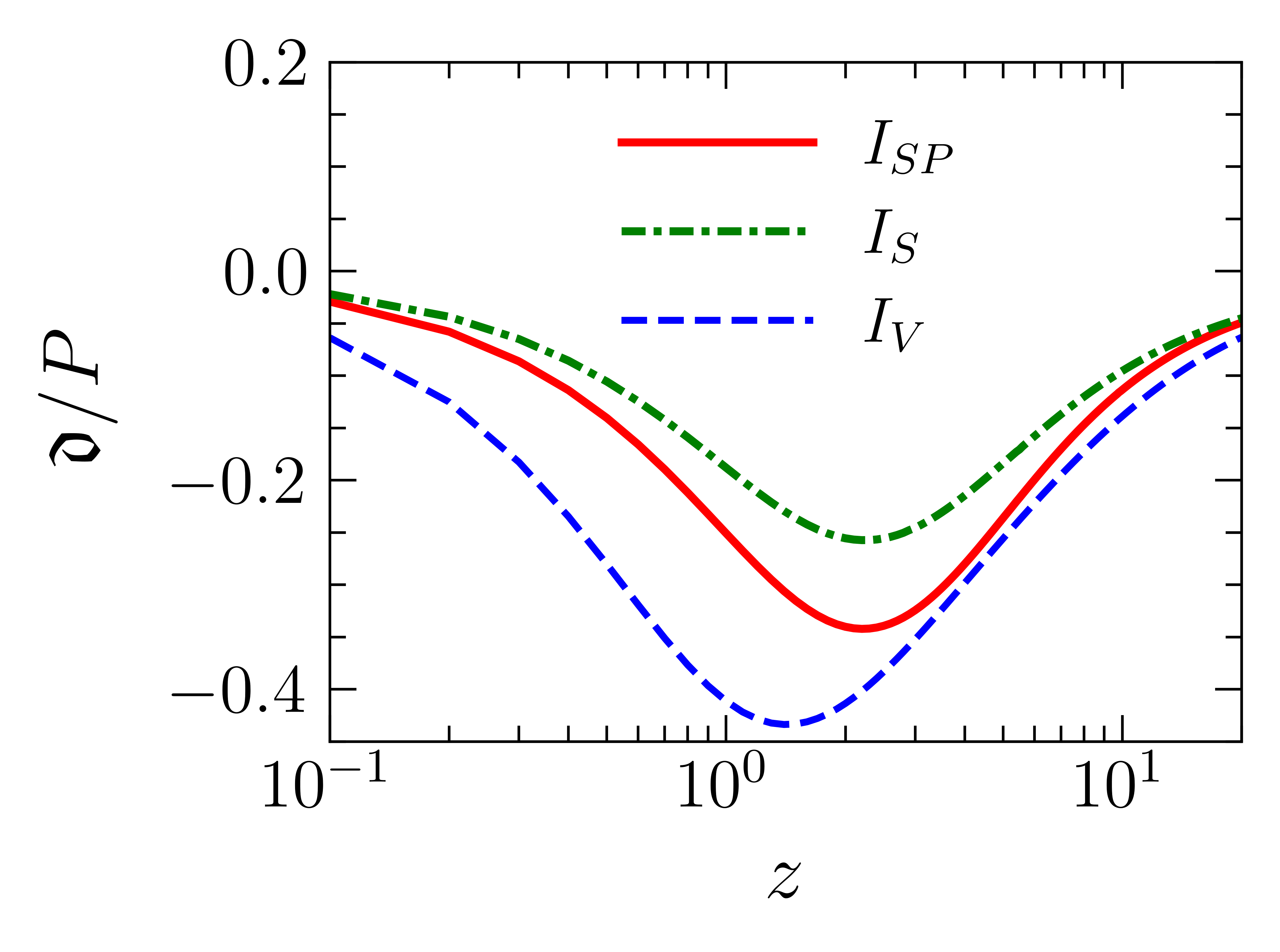}
	\end{minipage}
	\caption{Relaxation times ($\tauw$, $\tauk$, $\taut$) and the transport coefficient $\mathfrak{d}$ are shown as functions of $z=m/T$ for three different interactions.}
	\label{fig:spintranscoeff}
\end{figure*}
where $\eta>0$ and $\zeta >0$ denote the shear and bulk viscosity, respectively. It is well known that the first-order hydrodynamic theory is acausal and unstable \cite{Hiscock:1985zz}. This problem is removed by treating $\pi^{\mu\nu}$ and $\Pi$ as dynamical variables, the time evolution of which can be derived in the Denicol-Niemi-Molnár-Rischke (DNMR) framework as~\cite{Denicol:2012cn,Denicol:2015transcoeff}
\begin{align}
\dot{\Pi}&=\frac{\Pi_{\rm N S}-\Pi}{\tau_{\Pi}}-\frac{\delta_{\Pi \Pi}}{\tau_{\Pi}} \Pi\, \theta+\frac{\lambda_{\Pi \pi}}{\tau_{\Pi}} \pi^{\alpha\beta} \sigma_{\alpha\beta} \label{eq:Pi}\;,\\
\dot{\pi}^{\langle\alpha\beta\rangle}&=\frac{\pi_{\rm N S}^{\alpha \beta}-\pi^{\alpha\beta}}{\tau_\pi}
-\frac{\delta_{\pi \pi}}{\tau_\pi} \pi^{\alpha \beta} \theta+\frac{\lambda_{\pi \Pi}}{\tau_\pi} \Pi \,\sigma^{\alpha\beta}\nonumber \\
& \quad -\frac{\tau_{\pi \pi}}{\tau_\pi} \pi_\gamma^{\langle\alpha} \sigma^{\beta\rangle \gamma}+\frac{\varphi_7}{\tau_\pi} \pi_\gamma^{\langle\alpha} \pi^{\beta\rangle \gamma}\;\label{eq:pi}.
\end{align}
We consider the following expressions for the bulk and shear viscosities~\cite{Denicol:2015transcoeff,Denicol:2018wdp},
\begin{equation}
\eta  = C_\eta \frac{\varepsilon+P}{T} \;, \quad \zeta = 75\eta \left(\frac{1}{3}-c_s^2\right)^2\;,
\end{equation}
with the square of the speed of sound
\begin{equation}
c_s^2=\left.\frac{\partial P}{\partial \varepsilon}\right|_{n_B}
+\left. \frac{n_B}{\varepsilon+P} \frac{\partial p}{\partial n_B}\right|_{\varepsilon}\;.
\end{equation}
The coefficient $C_\eta$ is set equal to 0.12. The relaxation times for bulk and shear corrections as well as the second-order transport coefficients in Eqs.~\eqref{eq:Pi}-\eqref{eq:pi} are given by~\cite{Denicol:2015transcoeff,Transcoeff2014_p7}:
\begin{align}
&\tau_\pi = \frac{5\eta}{\varepsilon+P} \;,\quad \tau_\Pi = \frac{\zeta}{15\left(\frac{1}{3}-c_s^2\right)^2(\varepsilon+P)}\;,\nonumber\\
&\frac{\delta_{\Pi\Pi}}{\tau_\Pi} = \frac{2}{3}	\;, \quad  \frac{\lambda_{\Pi\pi}}{\tau_\Pi} = \frac{8}{5}\left(\frac{1}{3}-c_s^2\right)\;,\quad \frac{\delta_{\pi\pi}}{\tau_\pi} = \frac{4}{3}\;,\nonumber\\
& \frac{\lambda_{\pi\Pi}}{\tau_\pi} = \frac{6}{5}\;,\quad \frac{\tau_{\pi\pi}}{\tau_\pi} = \frac{10}{7}\;,\quad \frac{\varphi_7}{\tau_\pi} = \frac{9}{70 P \tau_\pi}\;.\label{eq:coef}
\end{align}
For the initial energy-momentum tensor, we use the model of Refs~\cite{Shen:2020jwv,Shen:2021lambda}, which is based on the Glauber collision geometry and local energy-momentum conservation, with the only difference being that we compute the thickness functions and wounded-nucleon densities using the optical limit of the Glauber model (cf. Appendix A of Ref.~\cite{Singh:2024cub} for details). The bulk viscous pressure ($\Pi$) and the shear-stress tensor ($\pi^{\mu\nu}$) are initialized to zero. In this work, we focus on the Au+Au collisions at the top RHIC energy of $\sqrt{s_{\rm NN}}=200$ GeV. The impact parameter $b$ corresponding to various centralities are taken from Ref.~\cite{Singh:2024cub}. The initial time for hydrodynamic evolution is fixed at $\tau_0=$1 fm.

A description of observables, such as momentum spectra and flow coefficients, does not require the evolution of spin equations, as these are decoupled from the background. The initial condition model parameters and the switching energy density for the particlization surface, where fluid cells are converted to hadrons, are estimated by fitting the model predictions to experimental data. This calibration has already been performed in Ref.~\cite{Singh:2024cub} for $\sqrt{s_{\rm NN}}=200$ GeV, and we use the same set of parameters in this work. The particlization surface is obtained using the \texttt{CORNELIUS} code~\cite{Huovinen2012}. At the particlization surface, we use a hadron sampler~\cite{Karpenko:2015xea,Schafer2022,samplerweblink} to generate particles from fluid elements. The resulting particle set serves as input to the SMASH transport model~\cite{SMASH2016prc,wergieluk_2024_10707746} for subsequent hadron interactions and decays. To validate the background hydrodynamic code, we rerun our simulations and reproduce the results of Ref.~\cite{Singh:2024cub} for three centrality classes, as shown in Fig.~\ref{fig:bulk}.

\section{Spin relaxation times and transport coefficients}
The spin relaxation times ($\tauw$, $\tauk$, $\taut$) and the first-order transport coefficient $\mathfrak{d}$, which have been computed in a \emph{Mathematica} notebook \cite{Wagner:2024fry,Mathematica,MERTIG1991345,SHTABOVENKO2016432,SHTABOVENKO2020107478}, are plotted as functions of $z=m/T$ in Fig.~\ref{fig:spintranscoeff} for three different interactions. All other transport coefficients in the evolution equations for the spin degrees of freedom $(\omega_0^\mu,\ \kappa_0^\mu,\ \mathfrak{t}^{\mu\nu})$ 
are discussed in Ref.~\cite{Wagner:2024fry}. 

\section{Definition of spatial average}
\begin{figure}[t]
	\includegraphics[scale=0.8]{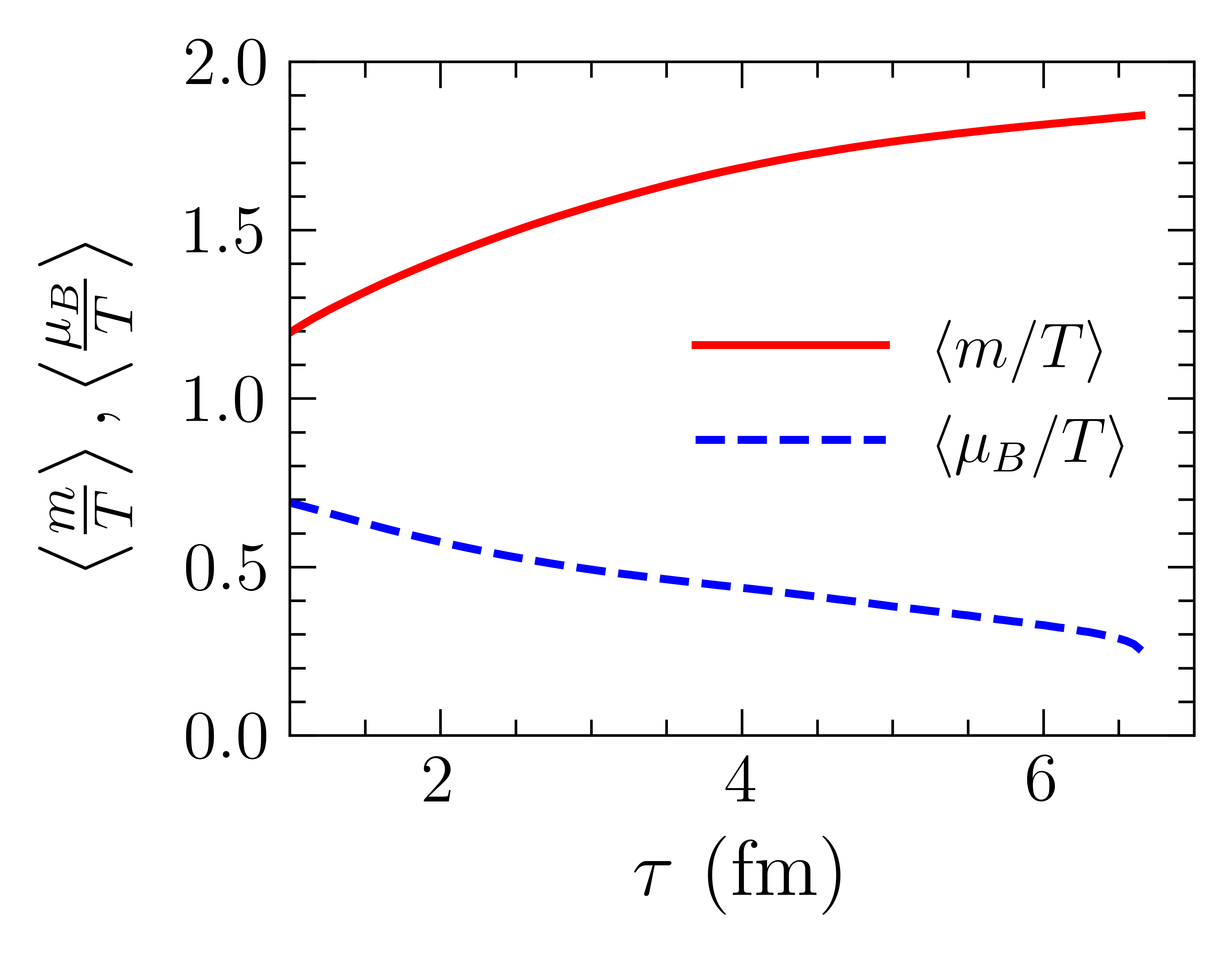}
	\caption{Time evolution of energy-weighted space-averaged $m/T$, for $m=300$ MeV, and $\mu_B/T$ for Au+Au collision at $\sqrt{s_{NN}}=200$ GeV, $b=8.4$ fm, and $\varepsilon_{\text{cut}}=0.5$ GeV/fm$^3$.}
	\label{fig:tempvst}
\end{figure}

We define the spatial average of a hydrodynamic quantity $A(\tau,x,y,\eta_s)$ as 
\begin{equation}
\label{eq:spaceavg}
\langle A\rangle (\tau) = \frac{\int \d x\ \d y\ \d\eta_s\ A(\tau,x,y,\eta_s)\varepsilon (\tau,x,y,\eta_s) \Theta (\varepsilon-\varepsilon_{\text{cut}})}{\int \d x\ \d y\ \d\eta_s\ \varepsilon (\tau,x,y,\eta_s) \Theta (\varepsilon-\varepsilon_{\text{cut}})}\;,
\end{equation}
where $\Theta$ is the Heaviside step function, and $\varepsilon_{\text{cut}}$ is a cutoff for fluid cells used to evaluate the integrals in the above expression. This prescription was used 
for the evolution of averaged spin potential in the main text with $\varepsilon_{\text{cut}}=0.5$ GeV/fm$^3$.
To better understand the evolution of the averaged spin potential, we show the time evolution of $\langle z\rangle = \langle m/T\rangle$, for $m=300$ MeV, and $ \langle \mu_B/T\rangle$ when $\varepsilon_{\text{cut}}=0.5$ GeV/fm$^3$ in Fig.~\ref{fig:tempvst}. The value of $\langle m/T\rangle$ increases from 1.2 to approximately 1.85, whereas $\langle \mu_B/T\rangle$ decreases from 0.7 to about 0.2 during the fireball's evolution. From Fig.~\ref{fig:spintranscoeff}, for $z$ in the range between 1.2 and 1.85, we observe that $\tauw$ is slightly larger than $\tauk$ for the $I_V$ and $I_{SP}$ interactions, whereas $\tauw$ is approximately four times larger than $\tauk$ for the $I_S$ interaction. Furthermore, $\tauw$ increases from approximately 2.5$\tau_\pi$ to 2.8$\tau_\pi$ for $I_V$, from about 2.0$\tau_\pi$ to 2.3$\tau_\pi$ for $I_{S}$, and from around 0.9$\tau_\pi$ to $\tau_\pi$ for $I_{SP}$. 
This behavior is reflected in the time evolution of the $xz$ component of the spin potential at late times, as demonstrated in the main text, where the relaxation of the spin potential toward the thermal vorticity proceeds fastest for the $I_{SP}$ interaction.

\begin{figure}[t]
	\includegraphics[scale=0.8]{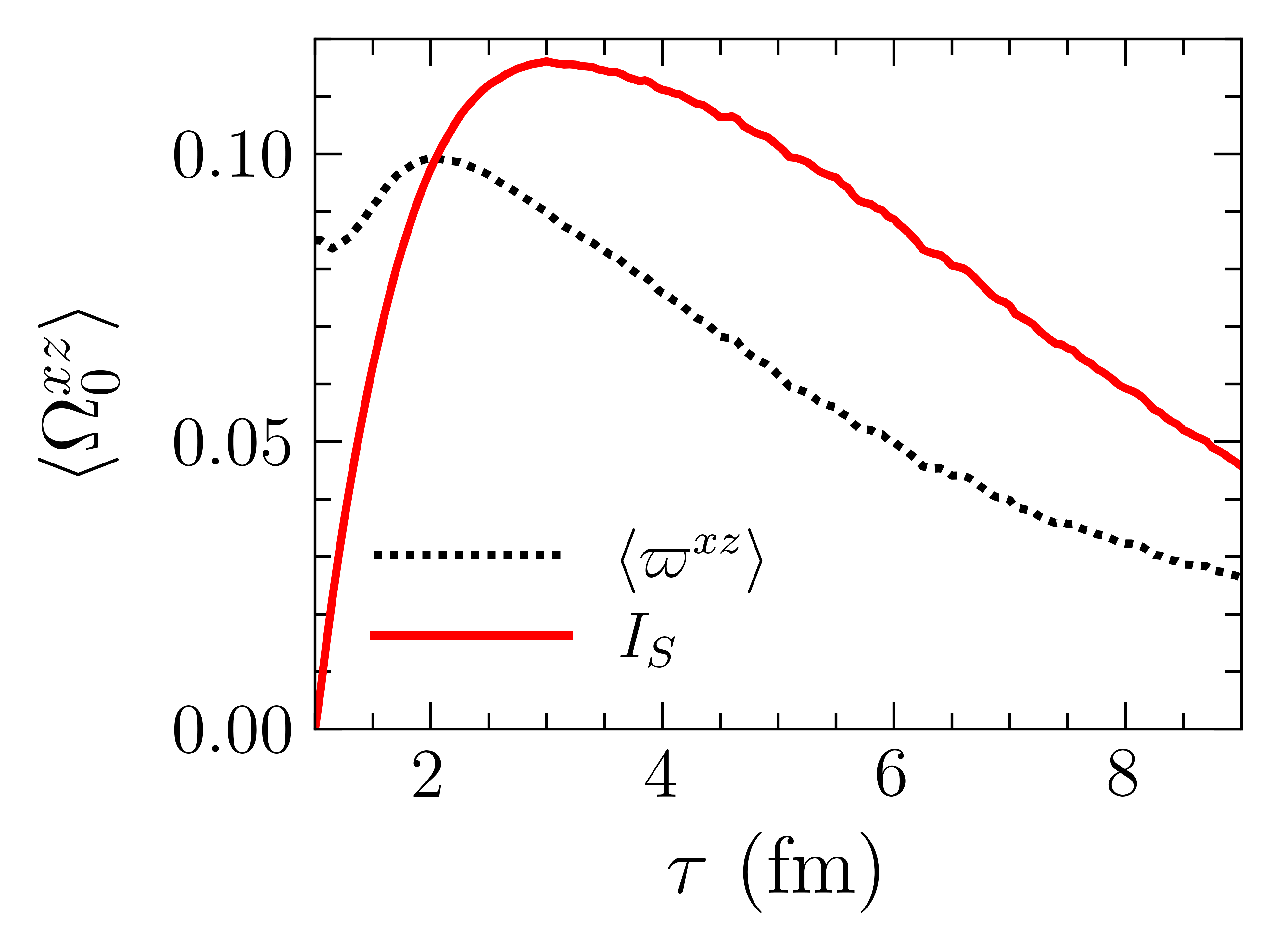}
	\caption{Time evolution of the energy-weighted space-averaged $xz$ component of the spin potential for scalar interaction and for $\varepsilon_{\text{cut}}=0.1$ GeV/fm$^3$.}
	\label{fig:spinpotevlncut0p1}
\end{figure}

In Fig.~\ref{fig:spinpotevlncut0p1}, we show the time evolution of $\langle \Omega_0^{xz}\rangle$ for a different choice of the cutoff, $\varepsilon_{\text{cut}}=0.1$ GeV/fm$^3$, for fluid cells in Eq.~\eqref{eq:spaceavg} which are used to compute the spatial average. A smaller cutoff means that the fluid cells with lower temperatures contribute to the spatial average. These low-temperature fluid cells will have longer relaxation times, leading to a delayed relaxation to thermal vorticity and a significant departure from equilibrium at the end of the fireball's lifetime. 

\section{\label{sec:numerics}Numerical details}
We are interested in the solution of equations having the form
\begin{equation}
(\partial_t + v^i\partial_i)\psi = -\frac{\psi -\psi_{\text{NS}}}{\tau_\psi} + S\;.
\label{eq:relaxadvection}
\end{equation}
\begin{figure*}[t]
	\includegraphics[scale=0.8]{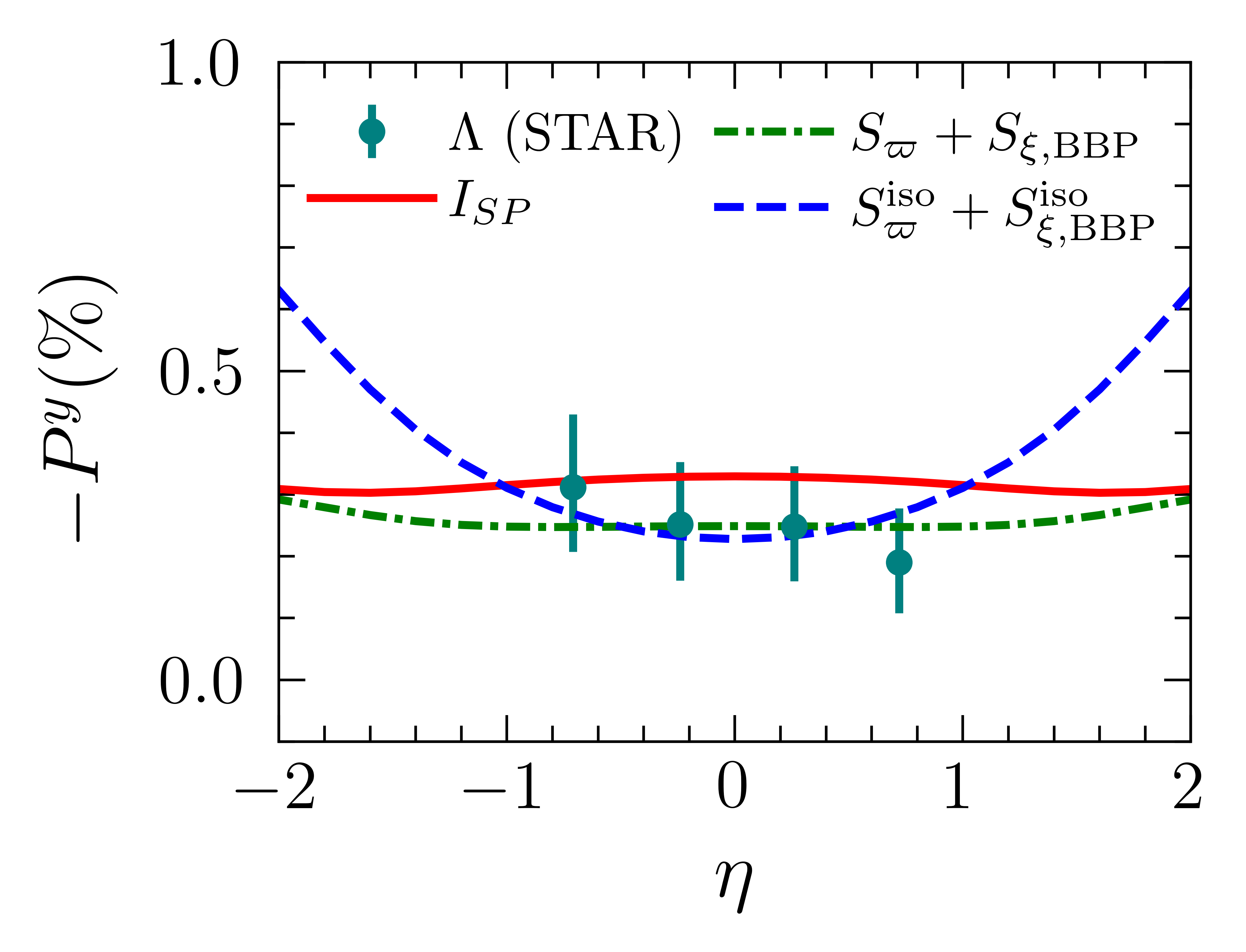}\hspace*{1in}
	\includegraphics[scale=0.8]{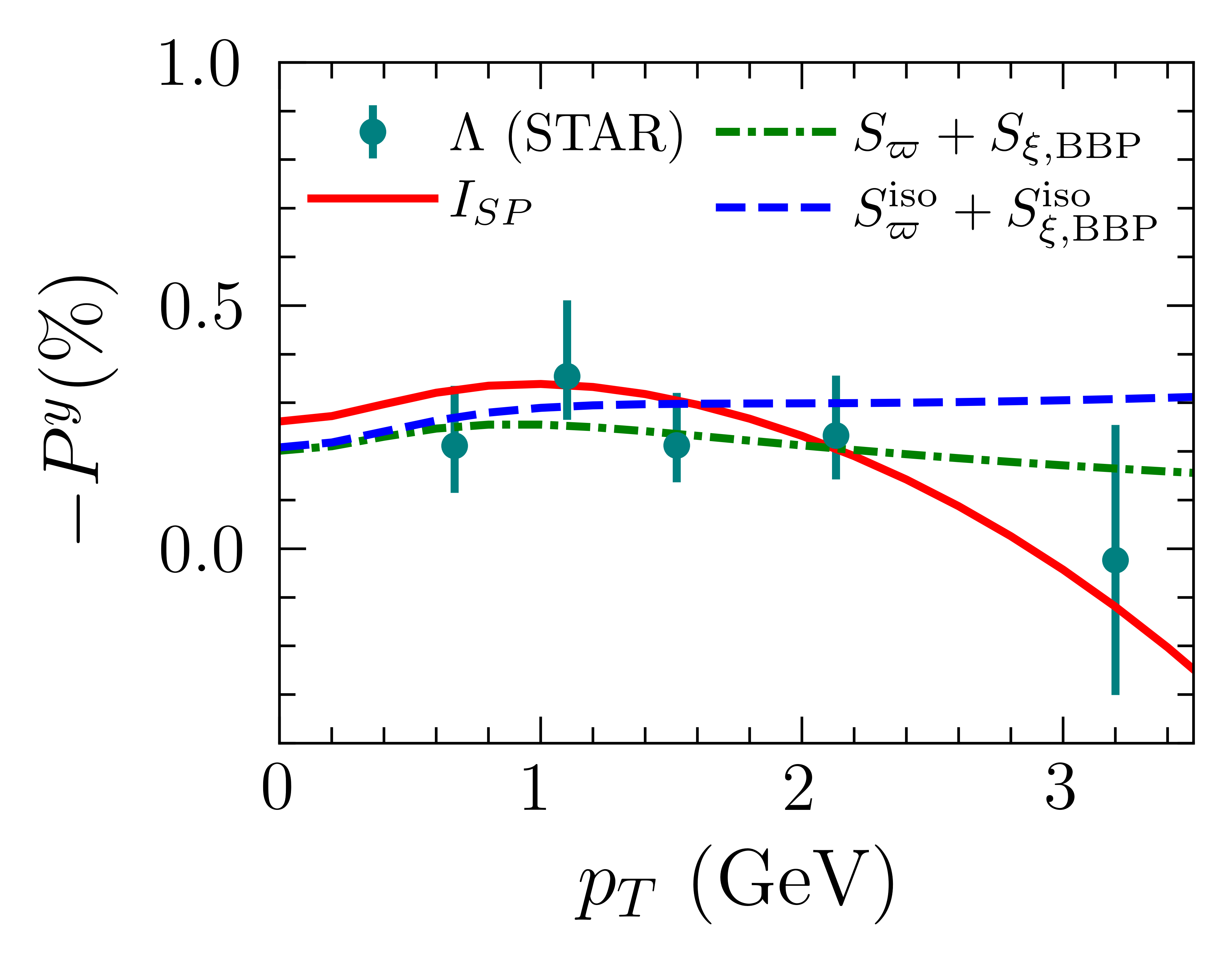}
	\hspace*{-0.1in}\includegraphics[scale=0.8]{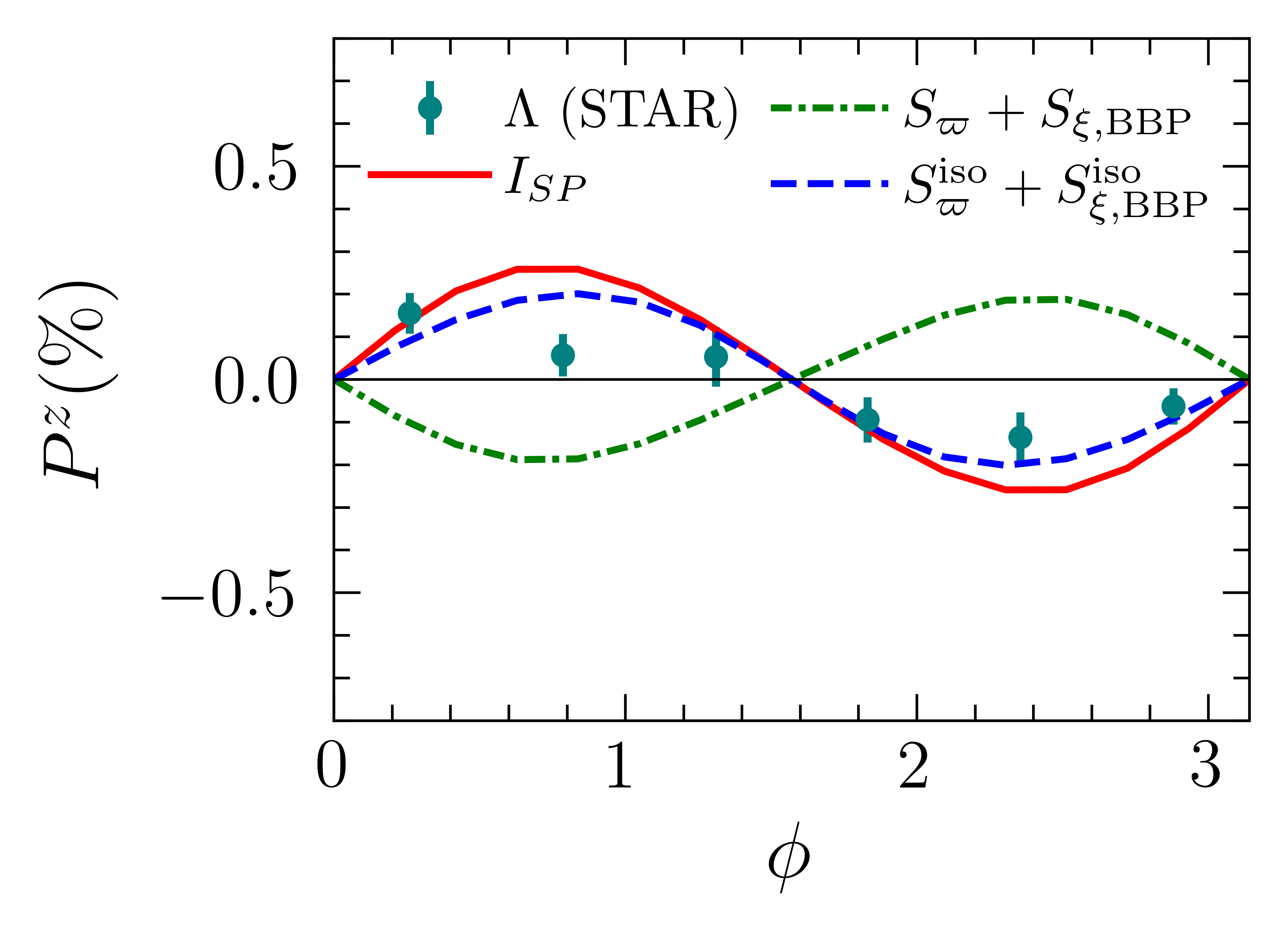}\hspace*{0.9in}
	\includegraphics[scale=0.8]{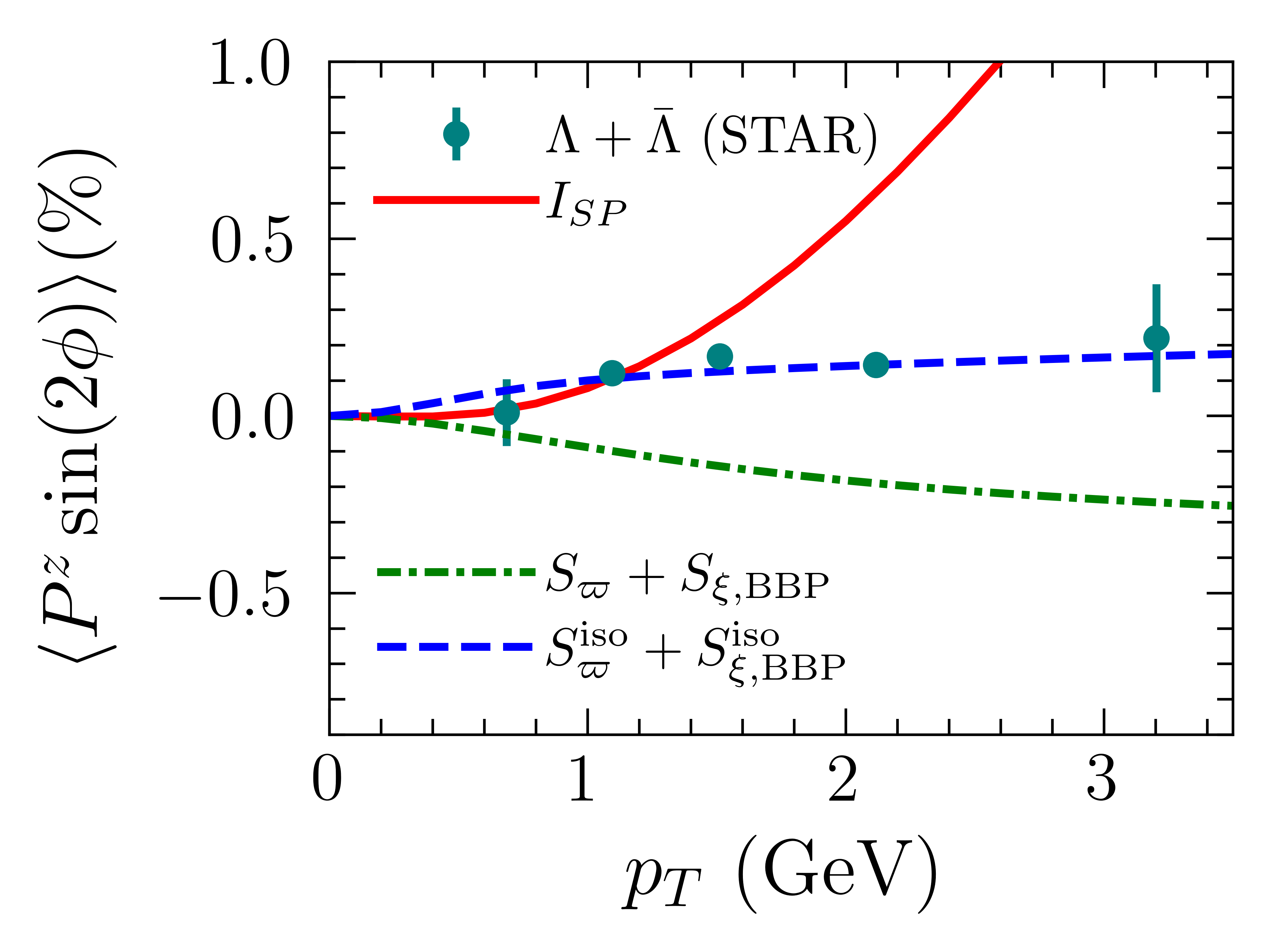}
	\caption{Comparison of results from dissipative spin hydrodynamics for $I_{SP}$ interaction scenario with BBP prescription for Au+Au collisions at $\sqrt{s_{NN}}=200$ GeV. The simulations are performed with fixed impact parameter, $b=8.4$ fm, corresponding to 20-50\% centrality. The top panel shows the (left) pseudorapidity and (right) transverse momentum dependence of the $y$ component of $\Lambda$ polarization. The bottom panel shows the (left) azimuthal and (right) transverse momentum dependence of the $z$ component of $\Lambda$ polarization. Experimental data are taken from Refs.~\cite{STAR200_Pj,STAR:2019erd} with the updated decay parameter ($\alpha_\Lambda = 0.732$).}
	\label{fig:compareBBP}
\end{figure*}
We first apply the Strang splitting method to split the problem into the following set of equations:
\begin{align}
\partial_t \psi &= -\frac{\psi -\psi_{\text{NS}}}{\tau_\psi} + S\;,\label{eq:relax}\\
(\partial_t + v^i\partial_i)\psi &= 0\;.\label{eq:advection}
\end{align}
We denote the value of $\psi$ at position $(x_i,y_j,\etas{}_{,k})$ and at $n^{\text{th}}$ time step by $\psi_{ijk}^n$, i.e., $\psi_{ijk}^n = \psi(t_n,x_i,y_j,\etas{}_{,k})$. We first solve Eq.~\eqref{eq:relax} according to the following prescription
\begin{widetext}
\begin{equation*}
   \psi _{ijk}^{\dagger n+1} = \left\{ \begin{array}{lcl}
\psi_{\text{NS}} + (\psi_{ijk}^n-\psi_{\text{NS}})\exp \left(-\frac{\Delta t}{\tau_\psi}\right)+\Delta t \ S(\psi_{ijk}^n) & , & \Delta t \le \tau_\psi \\
\psi_{ijk}^n - (\psi_{ijk}^n-\psi_{\text{NS}})\frac{\Delta t}{\tau_\psi}+\Delta t \ S(\psi_{ijk}^n) & , & \text{Otherwise}
\end{array}\right. \;.
\end{equation*}
\end{widetext}
We then use the solution above to solve Eq.~\eqref{eq:advection} using Corner Transport Upwind (CTU) method as follows~\cite{Karpenko:2013wva}:
\begin{equation*}
    \psi_{ijk}^{n+1} = \sum_{p,q,r=-1,0,1}P_p Q_q R_r\psi _{i+p,j+q,k+r}^{\dagger n+1}\;,
\end{equation*}
where $P_p$, $Q_q$ and $R_r$ denote the $p^{\text{th}}$, $q^{\text{th}}$ and $r^{\text{th}}$ elements of arrays $P$, $Q$ and $R$ given by
\begin{align*}
P &= \{-a_x^-,1-|a_x|,a_x^+\}\;,\\
Q &= \{-a_y^-,1-|a_y|,a_y^+\}\;,\\
R &= \{-a_{\etas}^-,1-|a_{\etas}|,a_\etas^+\}\;.
\end{align*}
Here $a_i^- = \min (v^i\Delta t/\Delta x^i,0)$ and $a_i^+ = \max (v^i\Delta t/\Delta x^i,0)$. The scheme outlined above is first order accurate in time. The scheme is made second order accurate using Predictor-Corrector method by repeating the procedure at an intermediate time-step ($t+\Delta t/2\equiv t^{n+1/2}$). The intermediate value of $\psi _{ijk}^{\dagger}$, for example, denoted by $\psi _{ijk}^{\dagger n+1/2}$ serves as the prediction, and is used to correct the solution $\psi _{ijk}^{\dagger n+1}$ by evaluating $S$ with the predicted value.

\begin{figure*}[t]
	\includegraphics[scale=0.85]{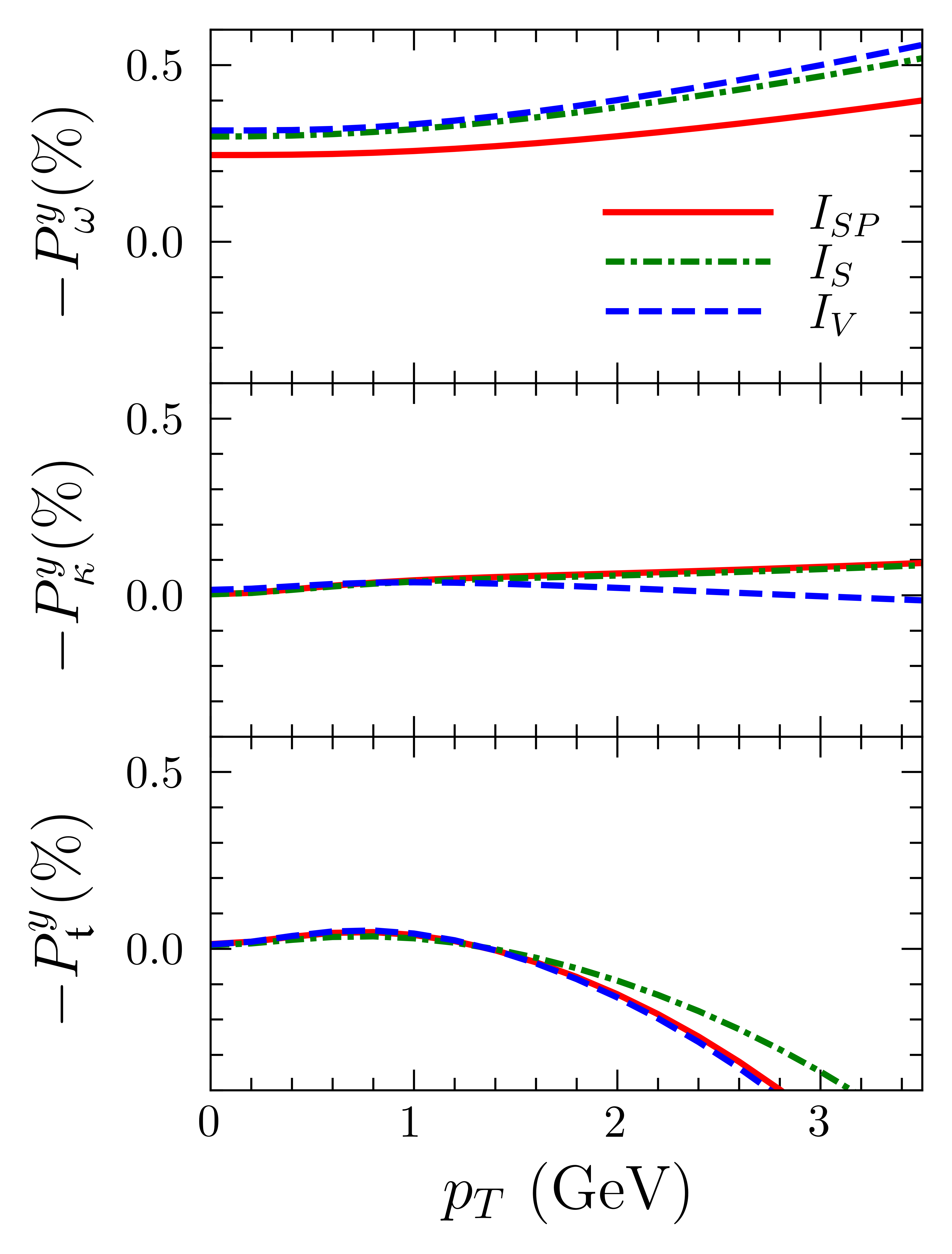}
	\includegraphics[scale=0.85]{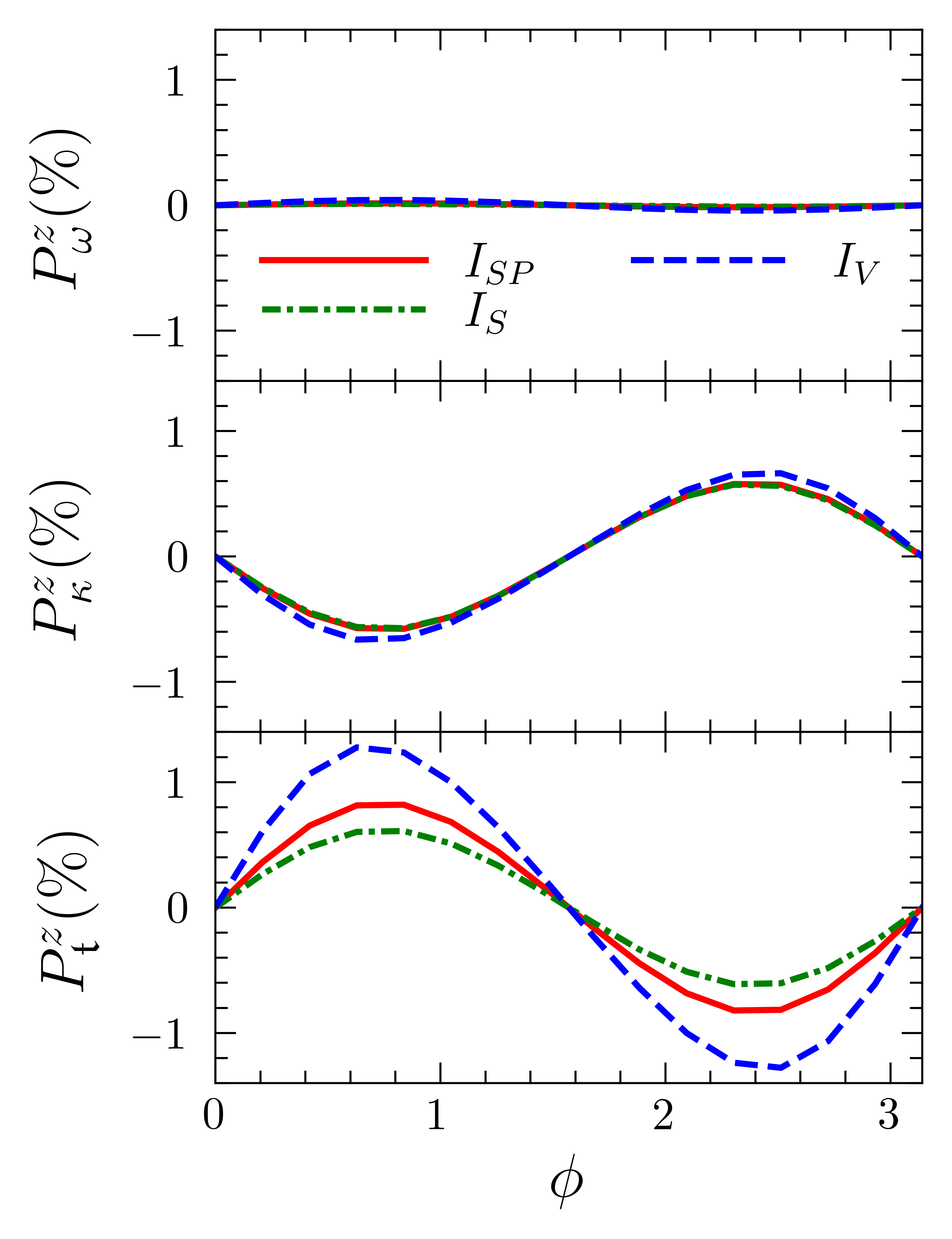}
	\caption{Individual contributions from $\omega_0^\mu$, $\kappa_0^\mu$ and $\mathfrak{t}^{\mu\nu}$ to the (left) $y$ component and (right) $z$ component of $\Lambda$ polarization for three different interactions. Note that $P^z_\omega$ is very small and appears as zero on the chosen scale of the plot, which is kept consistent to allow a visual comparison of the relative individual contributions.}
	\label{fig:spinindividual}
\end{figure*}

In order to cast 
the evolution equations for the spin degrees of freedom $(\omega_0^\mu,\ \kappa_0^\mu,\ \mathfrak{t}^{\mu\nu})$ in the form of Eq.~\eqref{eq:relaxadvection}, we follow the notation of Ref.~\cite{Karpenko:2013wva} and define a scaled four-vector, $\tilde{A}^\mu$, using a given four-vector $A^\mu$ as follows:
\begin{equation*}
\tilde{A}^\tau = A^\tau\quad ,\quad \tilde{A}^x=A^x \quad,\quad \tilde{A}^y=A^y\quad , \quad \tilde{A}^\etas = \tau A^\etas\;.
\end{equation*}
Similarly, a scaled rank-2 tensor is defined as
\begin{equation*}
\begin{array}{cccl}
\widetilde{S}^{\mu\nu} & = & S^{\mu\nu} & \text{ for }\mu,\nu \ne \etas\;,\\
\widetilde{S}^{\mu\nu} & = & \tau S^{\mu\nu} & \text{ if any one of } \ \mu, \nu \text{ is } \etas\;,\\
\widetilde{S}^{\etas\etas} & = & \tau^2 S^{\etas\etas}\;. &
\end{array}
\end{equation*}
We also denote the Navier-Stokes limits of the spin degrees of freedom as follows:
\begin{equation*}
    \otNS^\mu = -\frac{\oft_{\text{K}}^\mu}{T} \quad ,\quad \ktNS ^\mu = -\frac{\dot{\ut}^\mu}{T} \quad ,\quad \stNS^{\mu\nu} = \frac{\mathfrak{d}\sigt^{\mu\nu}}{T}\;.
\end{equation*}
The equations that we solve numerically are obtained from 
spin evolution equations after re-writing in the above notation and are given by:
\begin{align}
  \gt \left(\dt_\tau \ot^\mu +\vt^i\dt_i \ot^\mu \right) &=  \Rt^\mu_\omega - \ut^\mu \ot^\rho \ut^\alpha \Dt _\alpha \ut_\rho - I^\mu_{\omega,\text{G}}\;,\label{eq:eom_omega_scaled}\\
  \gt \left(\dt_\tau \kt^\mu +\vt^i\dt_i \kt^\mu \right) &=  \Rt^\mu_\kappa - \ut^\mu \kt^\rho \ut^\alpha\Dt _\alpha \ut_\rho - I^\mu_{\kappa,\text{G}}\;,\label{eq:eom_kappa_scaled}\\
  \gt \left(\dt_\tau \st^{\mu\nu} +\vt^i\dt_i \st^{\mu\nu} \right) &= \Rt^{\mu\nu}_{\mathfrak{t}} - \left[\st^{\mu \beta}\ut^{\nu}+\st^{\nu\beta}\ut^\mu\right] \ut^\rho \Dt_\rho \ut_\beta - I^{\mu\nu}_{\mathfrak{t},\text{G}}\;,\label{eq:eom_t_scaled}
\end{align}
for $i=x,y,\etas$. In the above equations, $\Rt^\mu_\omega$, $\Rt^\mu_\kappa$ and $\Rt^{\mu\nu}_\mathfrak{t}$ are given by the following expressions.
\begin{widetext}
\begin{align}
  \Rt^\mu_\omega &= -\frac{\ot^\mu-\otNS^\mu}{\tau_\omega} +\epsilon^{\mu\nu\alpha\beta}\ut_\nu \left[\lwk \Dt_\alpha \kt{}_{,\beta} - \kt{}_{,\beta}\ut^\gamma\Dt_\gamma\ut{}_{,\alpha}\right] +\dww \ot^\mu \theta + \left[\aww \ot{}_{,\nu} \sigt^{\mu \nu} +\awt \st^{\mu\nu}  \oft_{\text{K},\nu} \right]\;,\\
  \Rt^\mu_\kappa &= -\frac{\kt^\mu-\ktNS^\mu}{\tau_\kappa} +\epsilon^{\mu\nu\alpha\beta}\ut_\nu \left[\frac{1}{2}\Dt_\alpha\ot{}_{,\beta}+ \ot{}_{,\beta}\ut^\gamma \Dt_\gamma\ut_\alpha\right]+\dkk \kt^\mu \theta + \tkt \st^{\mu\nu} \ut^\alpha \Dt_\alpha \ut_\nu \nonumber \\
  &\qquad \qquad  +\left(\akk \sigt^{\mu\nu} +\frac{1}{2}\oft_{\text{K}}^{\mu\nu}\right)\kt{}_{,\nu} +\lkt \tilde{\Delta}^\mu_\lambda \tilde{\Delta}^\gamma_\nu \Dt_\gamma \st^{\nu\lambda}\;,\\
  \Rt^{\mu\nu}_{\mathfrak{t}} &= -\frac{\st^{\mu\nu}-\stNS^{\mu\nu}}{\tau_{\mathfrak{t}}} + \dtt \st^{\mu\nu}\theta +\att \st_\alpha ^{\langle \mu}\sigt^{\nu\rangle \alpha} +\frac{5}{3}\st_\alpha ^{\langle \mu}\oft_{\text{K}}^{\nu\rangle \alpha}+\ltk \Dt^{\langle\mu} \kt^{\nu\rangle}
  +\ttw \oft_{\text{K}}^{\langle\mu}\ot^{\nu\rangle}
  +\atw \sigt_\lambda{}^{\langle\mu}\epsilon^{\nu\rangle\lambda\alpha\beta} \ut_\alpha \ot{}_{,\beta}\;.
\end{align}
\end{widetext}

\begin{figure*}[t]
	\includegraphics[scale=0.8]{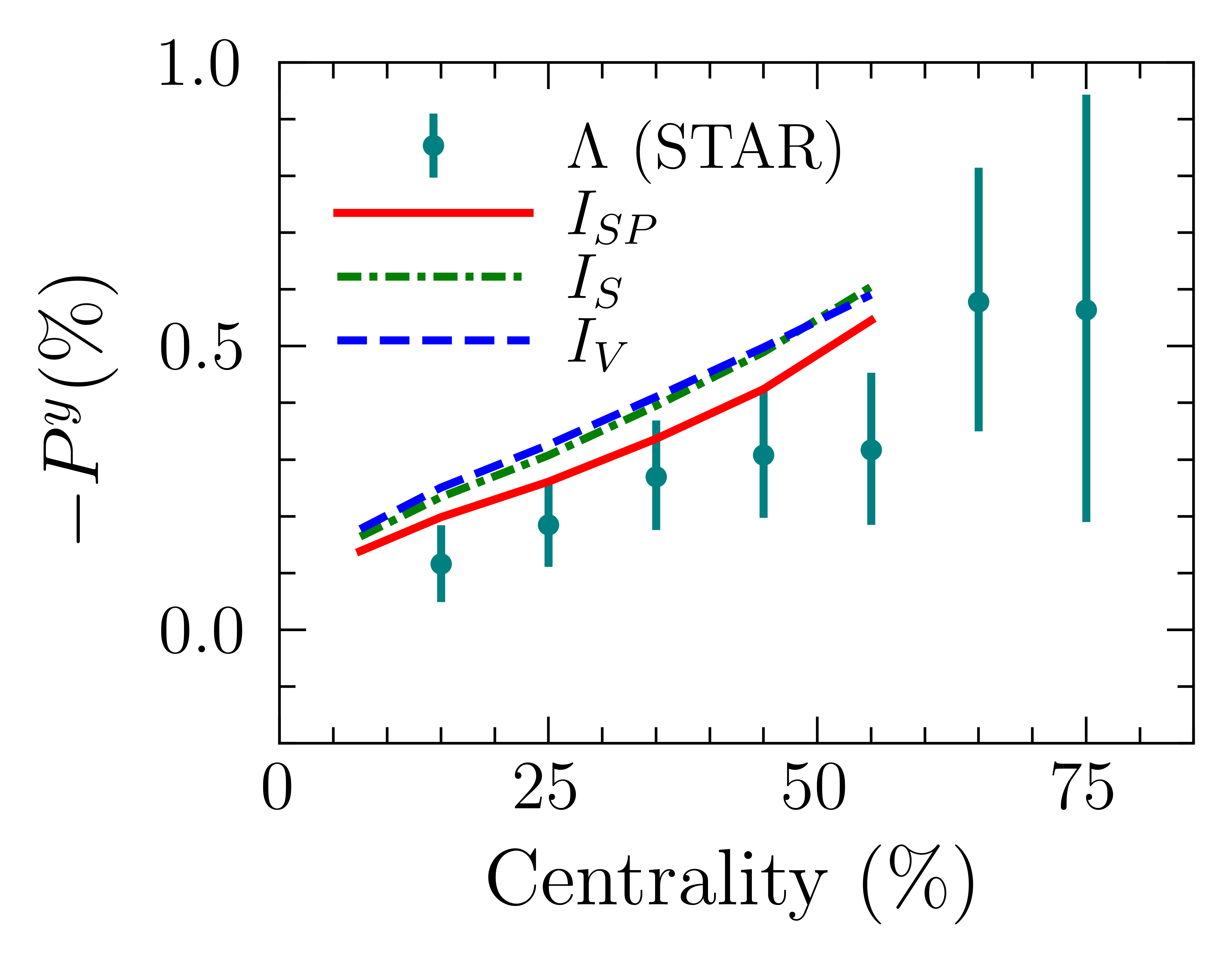}\hspace*{0.2in}
	\includegraphics[scale=0.8]{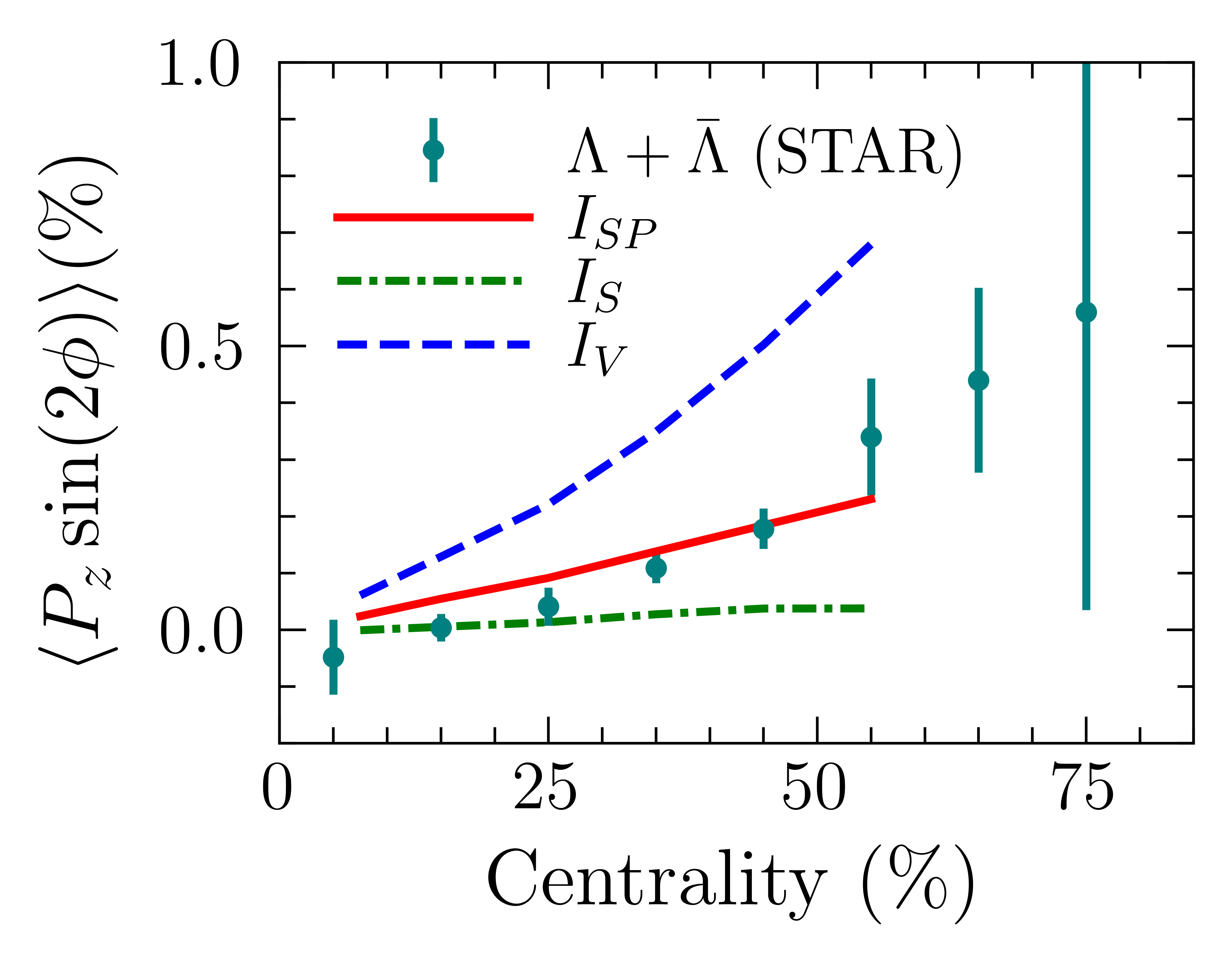}
	\caption{Comparison of results from dissipative spin hydrodynamics with experimental data on centrality dependence of $P^y$ and $P^z$ of $\Lambda$-spin polarization for three interaction scenarios. Experimental data are taken from Refs.~\cite{STAR200_Pj,STAR:2019erd} with the updated decay parameter ($\alpha_\Lambda = 0.732$).}
	\label{fig:polvscent}
\end{figure*}

The geometrical source terms in Eqs.~\eqref{eq:eom_omega_scaled}-\eqref{eq:eom_t_scaled}, $I^\mu_{\omega,\text{G}}$, $I^\mu_{\kappa,\text{G}}$ and $I^{\mu\nu}_{\mathfrak{t},\text{G}}$, arising from Christoffel symbols in the covariant derivatives are given by
\begin{align*}
  I^{\tau}_{\omega,\text{G}} &= \frac{\ut^\etas \ot^\etas}{\tau} \quad , \quad I^{\etas}_{\omega,\text{G}} = \frac{\ut^\etas \ot^\tau}{\tau} \quad , \quad I^{x}_{\omega,\text{G}} = I^{y}_{\omega,\text{G}} = 0\;,\\
  I^{\tau}_{\kappa,\text{G}} &= \frac{\ut^\etas \kt^\etas}{\tau} \quad , \quad I^{\etas}_{\kappa,\text{G}} = \frac{\ut^\etas \kt^\tau}{\tau} \quad , \quad I^{x}_{\kappa,\text{G}} = I^{y}_{\kappa,\text{G}} = 0\;,\\
  I^{\tau \tau}_{\mathfrak{t},\text{G}} &= \frac{2\ut ^\etas \st^{\tau\etas}}{\tau} \quad , \quad I^{\tau x}_{\mathfrak{t},\text{G}} = \frac{\ut ^\etas \st^{\etas x}}{\tau} \;,  \\
  I^{\tau y}_{\mathfrak{t},\text{G}} &= \frac{\ut ^\etas \st^{\etas y}}{\tau} \quad , \quad I^{\tau \etas}_{\mathfrak{t},\text{G}} = \frac{\ut ^\etas }{\tau}\left(\st^{\tau \tau} + \st^{\etas \etas}\right)\;, \\
  I^{\etas x}_{\mathfrak{t},\text{G}} &= \frac{\ut ^\etas \st^{\tau x}}{\tau} \quad , \quad I^{\etas y}_{\mathfrak{t},\text{G}} = \frac{\ut ^\etas \st^{\tau y}}{\tau} \;, \\
  I^{\etas \etas}_{\mathfrak{t},\text{G}} &= \frac{2\ut ^\etas \st^{\tau\etas}}{\tau} \quad , \quad I^{xx}_{\mathfrak{t},\text{G}} = I^{xy}_{\mathfrak{t},\text{G}} = I^{yy}_{\mathfrak{t},\text{G}} = 0\;.
\end{align*}

\section{Comparison to quantum-statistical approach}
We compare our results for $\Lambda$-polarization, calculated within our framework [using Eq.~(3) of main text],
with those obtained using the prescription from Refs.~\cite{Becattini:2015ska,Becattini:2021suc}, which is given by
\begin{equation}
\label{eq:bbp}
S^\mu(p) = S^\mu_{\varpi}(p) + S^\mu_{\xi,\text{BBP}}(p)\;,
\end{equation}
where $\hat{t}=(1,0,0,0)$,
\begin{subequations}
\label{eqs:components_S_BBP}
\begin{align}
S_{\varpi}^\mu(p) &= -\frac{1}{8m_\Lambda}\epsilon^{\mu\nu\rho \sigma}p_\sigma \frac{\int \d\Sigma\cdot p \ f_0\widetilde{f}_0 \varpi_{\nu\rho}}{\int \d\Sigma\cdot p \ f_0}\;,\\
S^\mu_{\xi,\text{BBP}}(p) &= -\frac{\epsilon^{\mu\nu\rho \sigma}}{4m_\Lambda} \frac{p_\sigma p^\lambda}{p\cdot \hat{t}}\frac{\int \d\Sigma\cdot p \ f_0\widetilde{f}_0 \hat{t}_\nu \xi_{\lambda\rho}}{\int \d\Sigma\cdot p \ f_0}\;.
\end{align}
\end{subequations}
Without any approximation, the longitudinal polarization due to the combined effects of thermal vorticity and thermal shear is negative, as shown by green dash-dotted lines in lower panel of Fig.~\ref{fig:compareBBP}. With the isothermal approximation \cite{Becattini:2021suc}, the surface $\Sigma$ is assumed to be at constant temperature (which is a good approximation at high energies), and factors of temperature are pulled out of the integral in the density operator rather than expanded in gradients. Under this approximation, defining $\Xi^{\mu\nu}=\frac12 \partial^{(\mu}u^{\nu)}$ and $\omega^{\mu\nu}= \frac12 \partial^{[\mu}u^{\nu]}$, the expressions \eqref{eqs:components_S_BBP} become
\begin{subequations}
\begin{align}
S_{\varpi}^{\text{iso},\mu}(p) &= -\frac{1}{8m_\Lambda}\epsilon^{\mu\nu\rho \sigma}p_\sigma \frac{\int \d\Sigma\cdot p \ f_0\widetilde{f}_0 \omega_{\nu\rho}}{T\int \d\Sigma\cdot p \ f_0}\;,\\
S^{\text{iso},\mu}_{\xi,\text{BBP}}(p) &= -\frac{\epsilon^{\mu\nu\rho \sigma}}{4m_\Lambda} \frac{p_\sigma p^\lambda}{p\cdot \hat{t}}\frac{\int \d\Sigma\cdot p \ f_0\widetilde{f}_0 \hat{t}_\nu \Xi_{\lambda\rho}}{T\int \d\Sigma\cdot p \ f_0}\;.
\end{align}
\end{subequations}
The combined effects of thermal vorticity and thermal shear in the isothermal approximation give the correct sign of longitudinal polarization, as shown by blue dashed lines in lower panel of Fig.~\ref{fig:compareBBP}. It is worth noting that the isothermal approximation employed in Ref. \cite{Becattini:2021suc}, if it were applied to the evolution equations for the spin degrees of freedom $(\omega_0^\mu,\ \kappa_0^\mu,\ \mathfrak{t}^{\mu\nu})$,
involves neglecting the temperature-gradient term, which would result in an asymptotic value of $\kappa_{0,\text{NS,iso}}^\mu =u_\nu \varpi^{\nu\mu}\approx -\dot{u}^\mu/(2T)$ -- half of the value obtained without the approximation.

\section{Additional figures}
In the left panel of Fig.~\ref{fig:spinindividual}, we show the individual contributions from $\omega^\mu_0$, $\kappa^\mu_0$ and $\mathfrak{t}^{\mu\nu}$ to the transverse momentum ($p_T$) dependence of the $y$ component of $\Lambda$ polarization for three different interactions, and in the right panel of Fig.~\ref{fig:spinindividual}, we show the individual contributions to the azimuthal angle dependence of the $z$ component of $\Lambda$ polarization. The net spin polarization is obtained by summing the individual contributions. 

In the main text, we fixed the impact parameter of Au+Au collisions to 8.4 fm, corresponding to the mean impact parameter in the 20–50\% centrality class for Au+Au collision at  $\sqrt{s_{NN}}=200$ GeV. In Fig.~\ref{fig:polvscent}, we show the centrality dependence of the $y$ and $z$ components of $\Lambda$-spin polarization. Results from dissipative spin hydrodynamics for the $I_{SP}$ interaction scenario agree well with experimental data.

\begin{figure}[t]
\includegraphics[scale=0.8]{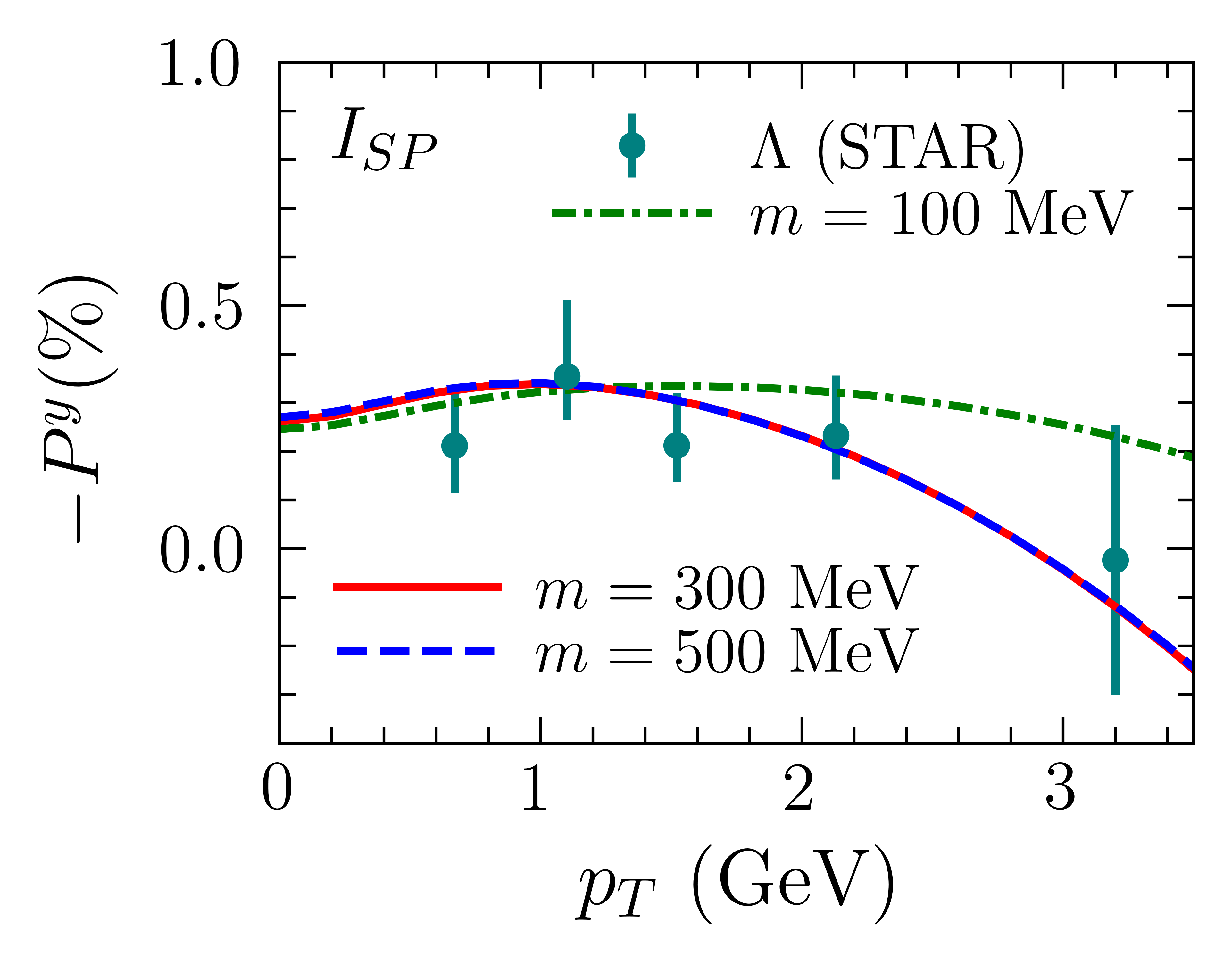}
\includegraphics[scale=0.8]{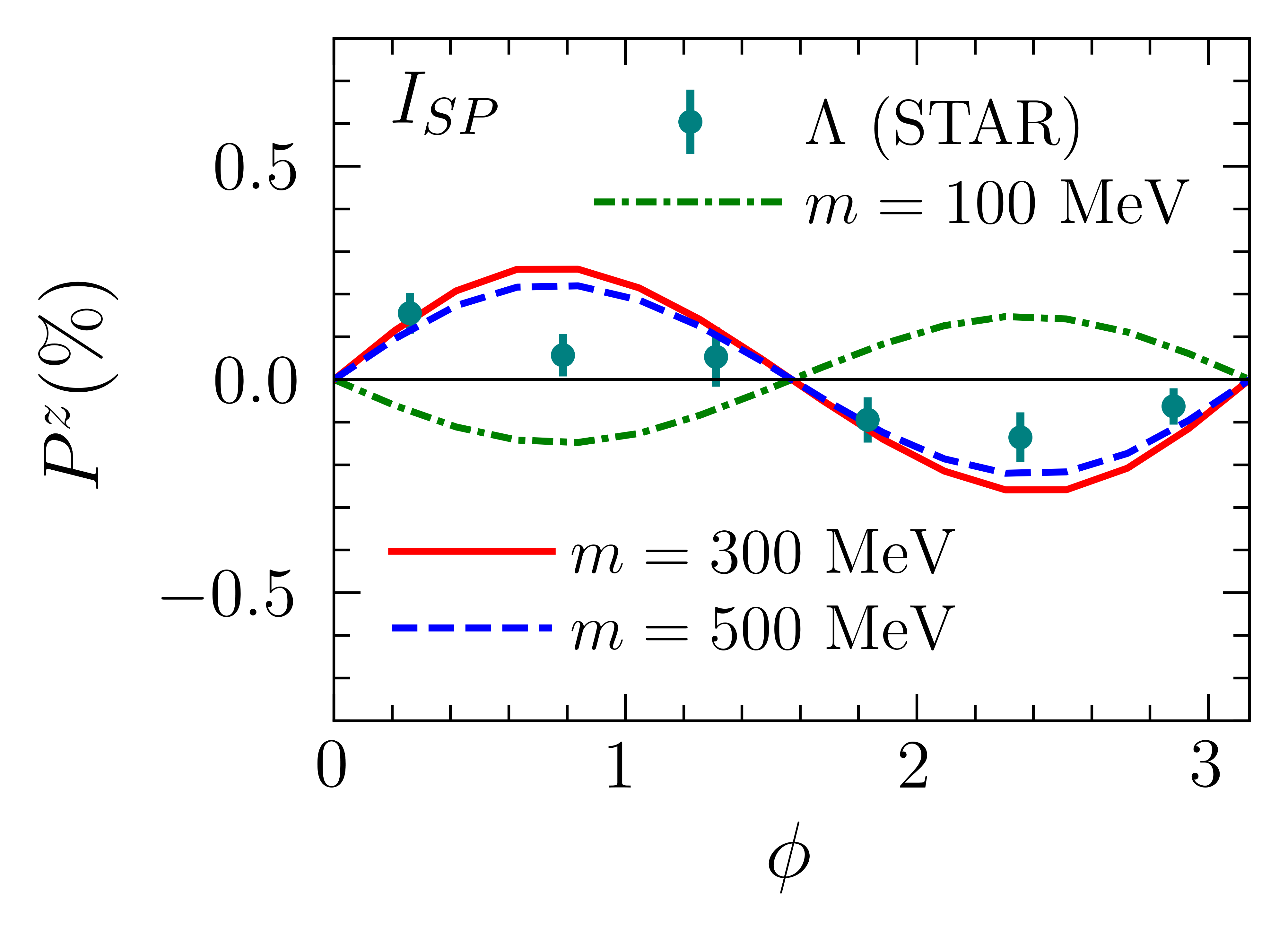}
\caption{For the $I_{SP}$ interaction scenario, the left panel shows the transverse momentum dependence of the 
$y$ component of polarization, while the right panel shows the azimuthal angle dependence of $z$ component for three different masses. Experimental data are taken from Refs.~\cite{STAR200_Pj,STAR:2019erd} with the updated decay parameter ($\alpha_\Lambda = 0.732$).}
\label{fig:mass}
\end{figure}

In Fig.~\ref{fig:mass}, we show the transverse momentum dependence of the $y$ component of polarization and the azimuthal angle dependence of the $z$ component of polarization for three different masses. The curves for masses of 300 and 500 MeV are nearly identical. In contrast, the sign of the longitudinal polarization is negative for $m=100$ MeV. These results can be understood by examining the $\mathfrak{d}$ vs. $z$ plot in Fig.~\ref{fig:spintranscoeff}. For $m=100$ MeV, $\langle z \rangle$ of the fireball varies between 0.4 and 0.62, with a smaller $\mathfrak{d}$, leading to reduced dissipative effects. For $m=500$ MeV, $\langle z \rangle$ varies between 2.0 and 3.1, resulting in a similar value of $\mathfrak{d}$, compared to the case of $m=300$ MeV.

\begin{figure}[H]
	\includegraphics[scale=0.8]{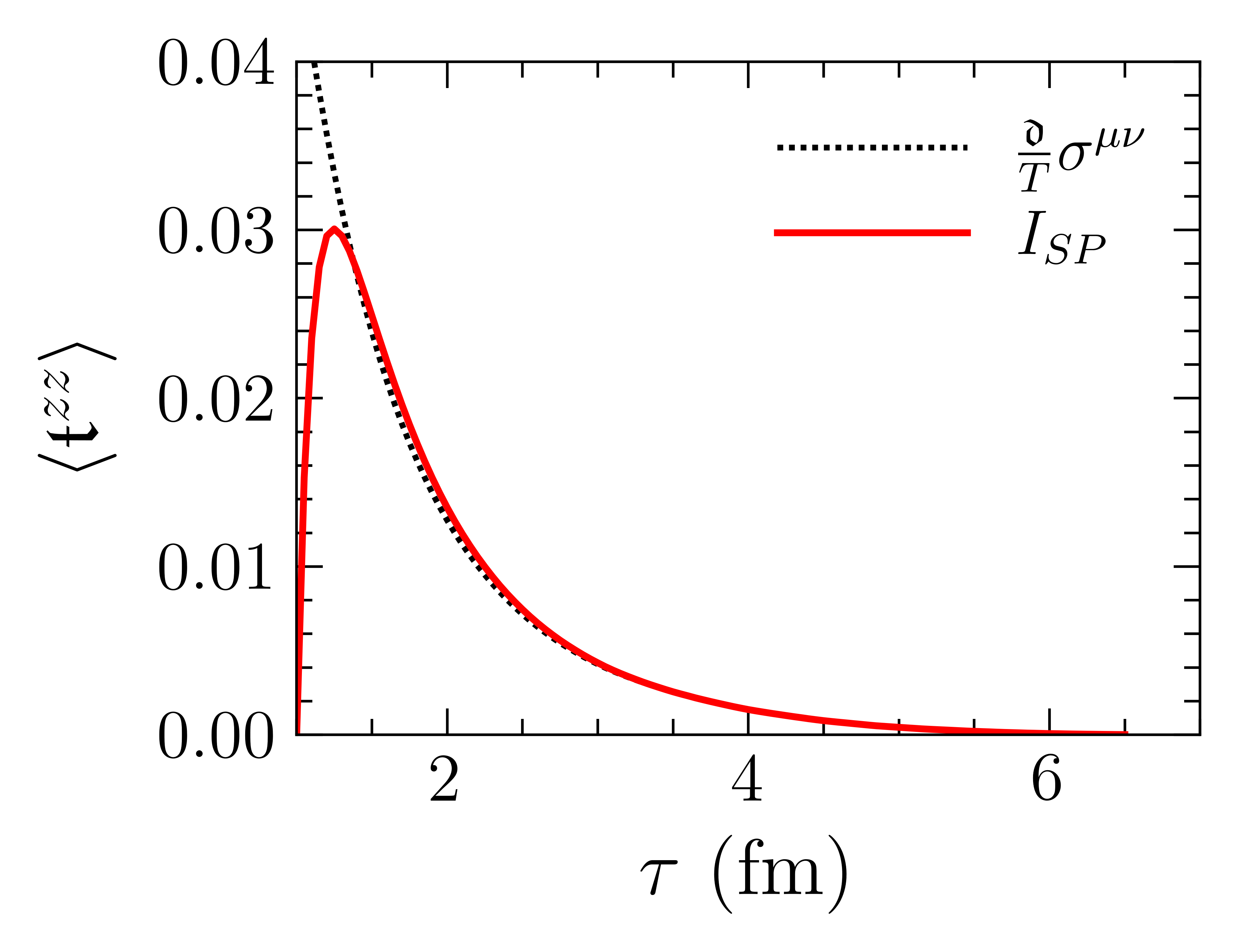}
	\caption{For the $I_{SP}$ interaction scenario, the spin-shear stress is shown to approach its asymptotic value.}
	\label{fig:spinshear_sp}
\end{figure}
In the main text, we demonstrated that the spin potential approaches its asymptotic value for all three interactions. However, for the spin-shear stress, we did not show the asymptotic values, as they differ for the three interactions due to the transport coefficient $\mathfrak{d}$. In Fig.~\ref{fig:spinshear_sp}, we demonstrate for the $I_{SP}$ scenario that the spin-shear stress does approach its asymptotic value $\frac{\mathfrak{d}}{T}\sigma^{\mu\nu}$.

%

\bibliographystyle{apsrev4-2}
\bibliography{spinhydro}

\end{document}